\documentclass[preprint,showpacs,preprintnumbers,amsmath,amssymb,showkeys,superscriptaddress]{revtex4}

\usepackage{graphicx}
\usepackage{hyperref}
\makeatother

\newcommand{\dib}{\centering\includegraphics[width=10 cm]} 
\newcommand{\dibuno}{\centering\includegraphics[width=9.5cm]} 
\newcommand{\beq}{\begin{equation}} \newcommand{\eeq}{\end{equation}}
\newcommand{\beqa}{\begin{eqnarray}}
\newcommand{\eeqa}{\end{eqnarray}}
\newcommand{\bneg}{\begin{bfseries}}
\newcommand{\eneg}{\end{bfseries}}
\newcommand{\biz}{\begin{flushleft}}
\newcommand{\eiz}{\end{flushleft}} 

\begin{document}

\title{The anatomy of the simplest Duflo-Zuker mass formula}

\author{Joel Mendoza-Temis} \affiliation{ {Instituto de Ciencias
    Nucleares, Universidad Nacional Aut\'{o}noma de M\'{e}xico, 04510
    M\'{e}xico, D.F., Mexico}}

\author{Jorge G.~Hirsch} \affiliation{ {Instituto de Ciencias
    Nucleares, Universidad Nacional Aut\'{o}noma de M\'{e}xico, 04510
    M\'{e}xico, D.F., Mexico}}

\author{Andr\'es P. Zuker} \affiliation{IPHC, IN2P3-CNRS, Universit\'e
  Louis Pasteur, F-67037 Strasbourg, France}


\begin{abstract}
  The simplest version of the Duflo-Zuker mass model (due entirely to
  the late Jean Duflo) is described by following step by step the
  published computer code. The model contains six macroscopic monopole
  terms leading asymptotically to a Liquid Drop form, three
  microscopic terms supposed to mock configuration mixing (multipole)
  corrections to the monopole shell effects, and one term in charge of
  detecting deformed nuclei and calculating their masses. A careful
  analysis of the model suggests a program of future developments that
  includes a complementary approach to masses based on an
  independently determined monopole Hamiltonian, a better description of
  deformations and specific suggestions for the treatment of three
  body forces.

\end{abstract}
\maketitle

\section{Introduction}\label{sec:intro}
Masses are a fundamental property of nuclei, whose accurate knowledge
is important for a large number of processes in nuclear physics, in
particular in astrophysical phenomena~\cite{Rol88}, and for a variety
of applications in other areas, from elementary particle physics to a
precise determination of the kilogram.  Though much progress has been
made in measuring the masses of exotic nuclei (see for example
Ref.\cite{Bla06} and references therein), theoretical models are still
necessary to {\em predict} them in regions far from
stability~\cite{Lunn03}.  Advances in the calculation of atomic masses
have been hampered by the absence of a full theory of the nuclear
interaction and by the difficulties inherent to quantum many-body
calculations.  There has been much work in developing mass formulas
with either microscopic and macroscopic input or within
 a fully microscopic framework.

Soon after the latest compilation of nuclear masses AME03 \cite{AME03}
was published, a comparison between the predictions of a large set of
mass models, which were fitted to describe the nuclear masses included
in AME95 \cite{AME95}, for the more than 300 new masses included in
AME03 was presented \cite{Lunn03}. It became evident that the
predictions of the Duflo-Zuker (DZ) model~\cite{Duf95} were
outstanding. Recent tests of the predictive power of nuclear models
\cite{Men08} confirm the ability of DZ to make stable predictions
which are more accurate than those offered by other models.

The model has many parameters: 28 in the original version, tabulated
by Audi~\cite{dzcode} and used in the comparisons~\cite{Lunn03}. The
code is unpublished but a version with up to 33 parameters circulates
and it is now available in~\cite{dzcode}. Another code, with only 10
parameters (DZ10)~\cite{dzcode} leads to good though less spectacular
results but embodies the essence of the model.

Our task is to examine in full detail what DZ10 does and as we go
point to possible ways to improve on it, based on what has been
learned in recent years. To anticipate on what follows: Of the ten
terms enumerated in the abstract the three---crucial ones---supposed
to represent configuration mixing are anomalous in the sense that they
contain three body effects of unknown origin and they scale with the
total number of particles $A$ instead of $A^{1/3}$ as shell effects
should. They will be examined in special detail.

In Section~\ref{sec:basic} the basic elements of the DZ10 model are described.
Section~\ref{sec:fits} deals with the evolution of the results as the ten
parameters of the fits are switched on; the origin of the anomalous
scaling becomes apparent. Predictive power and stability of the model are
examined in Section~\ref{sec:predictive}. An analysis of the action of the anomalous
terms follows in Section~\ref{sec:ano2}. A full explanation of their nature is
given in Section~\ref{sec:macro} by comparing the macroscopic terms with an
independently determined monopole Hamiltonian. Section~\ref{sec:3b} is devoted
to the need to introduce three body forces. Section~\ref{sec:homage} is the
conclusion and an homage to Jean Duflo.

\section{Basic elements}\label{sec:basic}

\begin{figure}[h]    
		\dibuno{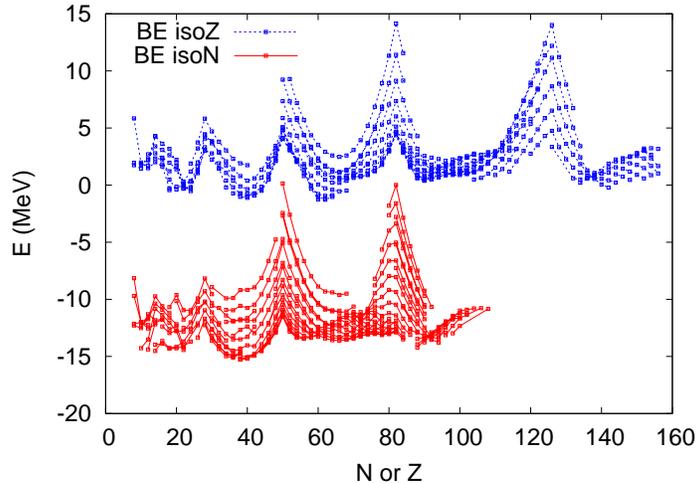}
    \caption{Shell effects (BE(exp)-E(LD)) along isotope and isotone
      lines (latter displaced by -14 MeV). Only even-even nuclei
        are shown.}\label{fig:beld}
\end{figure}

To illustrate what a model of nuclear masses is supposed to do
consider Fig.\ref{fig:beld} where experimental binding energies of
even-even nuclei are subtracted from an improved version of the
Bethe-Weizs{\"a}cker liquid drop (LD) form, given in Eq.(\ref{ld}) for
$Z$ protons, and $N$ neutrons, with mass number $A=N+Z$ and isospin
$T=\left|N-Z\right|/2$. The asymmetry term is generalized to include
Wigner and surface contributions.  Only even-even nuclei are shown as
they contain all the basic information. The pairing term is omitted
leading to (mostly) positive definite shell effects.
\begin{gather}
  E(LD)=15.5 A-17.8
A^{2/3}-28.6\frac{4T(T+1)}{A}+40.2\frac{4T(T+1)}{A^{4/3}}
-\frac{.7Z(Z-1)}{A^{1/3}}.
\label{ld} 
\end{gather}
 Both graphs contain exactly the same information. Four
remarks (injunctions):
\begin{itemize}
\item {\bf LD} Any model must contain the LD either explicitly or
  asymptotically.
\item {\bf EI} The only observed closures worth considering are the
  ``extruder-intruder'' (EI) ones at $N,Z$= (14),28,50,82 and 126, in
  which the subshell of largest angular momentum is ``extruded''
  from harmonic oscillator (HO) shell of principal quantum number $p$
  and becomes the intruder in the $p-1$ HO shell.
\item {\bf SPH-DEF} Flat patterns at the bottom of Fig.\ref{fig:beld},
  as in the regions 90 $<$ N $<$ 110 and 60 $<$ Z $<$ 75, correspond
  to well deformed nuclei. Any model must treat them specifically.
\item ${\bf A^{1/3}}$. Shell effects---defined by subtracting any
  contribution to the total binding energy from its asymptotic
  form---should scale as $A^{1/3}$.
 \end{itemize}

 These injunctions were incorporated in steps. First came the purely
 phenomenological work of Duflo~\cite{Duf94}. It was based on simple
 algebraic forms that fitted masses with RMSD of some 350
 keV. However, it extrapolated poorly. In a companion
 paper~\cite{Zuk94} it was shown that the algebraic simplicity could
 be reconciled with a microscopic derivation provided the microscopic
 shell effects be separated from the macroscopic (LD) form.  Then the
 observed closures were taken to define conventional shell model
 spaces and it was shown that configuration mixing would lead to
 simple quadratic and quartic forms in the number of valence
 particles~\cite{Zuk94}. The trouble was that the observed closures
 had to be put by hand.  The breakthrough came by incorporating hints
 from~\cite{Dufour} showing that from realistic interactions one could
 extract a ``master term'' containing both the leading LD bulk energy
 and strong harmonic oscillator (HO) shell effects. It was left to
 shift the HO closures to the observed ones and then use the shell
 model forms derived in~\cite{Zuk94} to mock configuration mixing. In
 a recent paper~\cite{Zuk08} the DZ strategy is presented in some
 detail, stressing in particular that it is not a ``mass formula'' but
 a functional of the orbital occupancies and explaining its success in
 dealing with well deformed nuclei. Our aim here is to show how to
 incorporate the last injunction above, by unearthing the reasons why
 DZ fails to respect it, which we (cryptically) anticipate with the
 benefit of hindsight: DZ assumed that asymptotic behaviour involves
 an LD form which ignores $A^{1/3}$ terms that turned out to be
 present and treated as genuine shell effects. Next section starts
 dealing with the problem.

\subsection{The Master term}\label{sec:master}
In DZ~\cite{Duf95} it was assumed (guessed) that realistic two body
interactions generate two collective terms solely responsible for the
leading LD contributions, of the form
\begin{gather}
\label{master}
M_A= \frac{\hbar \omega}{\hbar \omega_0} \left(\sum_p
  \frac {m_p}{\sqrt{D_p}}\right)^2, \, \, \,
M_T=  \frac{\hbar \omega}{\hbar \omega_0} \left(\sum_p
  \frac {t_p}{\sqrt{D_p}}\right)^2
\end{gather}
where $\hbar \omega$ is the HO frequency, $\hbar \omega_0$ is left as
a free  parameter, $D_p=(p+1)(p+2)$ is the degeneracy of
the major HO shell of principal quantum number $p$, $m_p=n_p+z_p, \;
t_p=n_p-z_p$, where $n_p,z_p$ are number operators for neutrons and
protons, respectively.

To obtain asymptotic estimates for $M_A$ we assume at first $N=Z$ and,
following Ref.~\cite{Bohr98}, sum up to the closed Fermi shell $p_f$,
which we associate to the total number of particles $A$ through (use
$p^{(3)}=p(p-1)(p-2)$; $\Longrightarrow$ leads to, implies;
$\approx$ approximately;  $\therefore$ therefore)
\begin{gather}
\label{A}
A=\sum_p m_p=\sum_{p=0}^{p_f}2D_p=\frac {2(p_f+3)^{(3)}}{3} 
\approx \frac {2(p_f+2)^{3}}{3} 
\Longrightarrow (p_f+2) \approx(\frac{3}{2}A)^{1/3}
\end{gather}
Next we estimate $\hbar\omega$ by relating it to the observed square
radius~ \cite{Bohr98},
\begin{gather} 
  \langle r^2 \rangle=\frac {\hbar}{A M_{nucl}\omega}\sum_p m_p(p+3/2)
  \approx \frac{3\hbar}{4M_{nucl}\omega}(p_f+2) \therefore
  \;\;\hbar\omega= 35.59\frac{A^{1/3}}{\langle r^2 \rangle} {\rm
    MeV}.\label{hw.scale}
\end{gather}
where $M_{nucl}$ is the nucleon mass.  The 35.59 coefficient
follows form an accurate estimate of the nuclear radii of magic and
semimagic nuclei~\cite{DZ02}, which may serve as a reminder of the
need to incorporate an isospin dependence,
 \begin{equation}
\label{R}
 \sqrt{\langle r_{\nu\pi}^2\rangle}\approx
A^{1/3}\left(.943-0.4\frac{t}{A^{4/3}}-
0.34(\frac{t}{A})^2\right)e^{(1.04/A)}; \quad t=N-Z.     
\end{equation}
In what follows, whenever convenient, we shall replace $\hbar
\omega/\hbar \omega_0$ or $\hbar \omega$ by the scaling factor
$1/\rho$ suggested by Eq~(\ref{hw.scale}), where
\beq \rho= \langle r^2\rangle/A^{1/3}
=A^{1/3}\left[1-0.5\left(\frac{t}{A}\right)^{2}\right]^{2} .
\label{rc}
\eeq
retaining the DZ10 choice for $ \langle r^2\rangle$
rather than the more sophisticated Eq.~(\ref{R}).

Approximating $\sqrt{D_p}\approx p+3/2$, leads to
\begin{gather}
  \sum_{p}^{p_f} {\frac {m_p}{\sqrt{D_p}}} =
  \sum_p^{p_f} 2\sqrt{D_p} \approx p_f(p_f+4) 
\approx (p_f+2)^2\label{MA1}\\
  M_A= \frac{1}{\rho}\left(\sum_p \frac {m_p}{\sqrt{D_p}}\right)^2
  \approx \frac{1}{\rho} (p_f+2)^4
\label{MA2}
\end{gather}
Similarly, for $N$ and $Z$ separately ($\equiv$ equivalent)
\begin{gather}
N \approx \frac {(p_{f\nu}+2)^{3}}{3}, ~~~~Z\approx \frac {(p_{f\pi}+2)^{3}}{3}, \\
M_N \equiv \frac{1}{\rho} \left(\sum_p^{p_{f\nu}} \frac {n_p}{\sqrt{D_p}}\right)^2  
\approx \frac {(p_{f\nu}+2)^4}{4\rho} 
\approx  \frac {(3N)^{4/3}} {4A^{1/3}}, 
\\
M_Z  \equiv \frac{1}{\rho} \left(\sum_p^{p_{f\pi}} 
\frac {z_p}{\sqrt{D_p}}\right)^2  
\approx  \frac {(p_{f\pi}+2)^4}{4\rho} 
\approx  \frac {(3Z)^{4/3}} {4A^{1/3}},
\end{gather}
where $p_{f\nu}, ~p_{f\pi}$ are the neutron and proton Fermi levels. As
\begin{equation}
M_A = M_Z + M_P + 2 \sqrt{M_Z} \sqrt{M_P}, ~~~
M_T = M_Z + M_P - 2 \sqrt{M_Z} \sqrt{M_P},
\end{equation}
the leading asymptotic ($\asymp$) estimates in  $t/A$ become
\begin{equation}
 M_A\asymp \left({\frac 3 2}\right)^{4/3} A \left(1-{\frac 2 9} \left({\frac t A}\right)^2\right); \qquad 
 M_T\asymp \left({\frac 3 2}\right )^{4/3} A {\left( \frac {2t} {3A}\right)}^2
\label{asymt} 
\end{equation}
To go beyond leading order is delicate, but combining Eqs.~(\ref{A})
and~(\ref{MA1}) we find that $A^{4/3}$ goes as
$p_f^4+8p_f^3+23p_f^2+\ldots$ against $p_f^4+8p_f^3+16p_f^2$ for
$(\sum_p m_pD_p^{-1/2})^2$, which points to a substantial
$A^{1/3}\,(\approx p_f+2)$ contribution to $M_A$ but a vanishing one
in $A^{2/3}$. In other words---as it stands---the master term has no
contribution to the LD surface energy. To look for its possible origin
we examine the microscopic derivation of $M_A$.

\subsection{Origin of the master terms. Scaling}\label{sec:orig}
\begin{figure}[t]
  \begin{center}
  \includegraphics[width=9 cm]{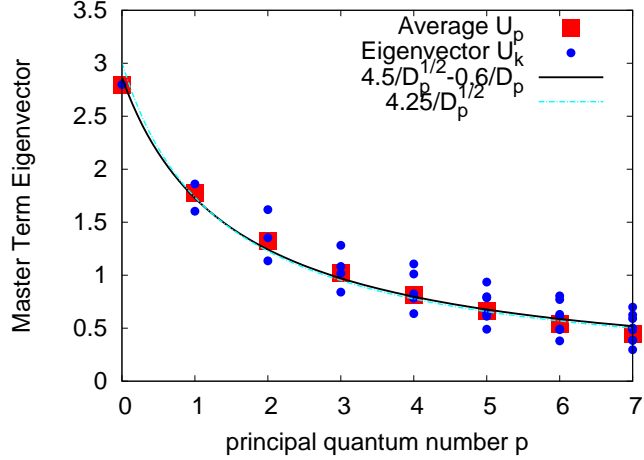}
  \caption{The isoscalar master eigenvector (boosted by an arbitrary
    factor) for the N3LO interaction. According to Eq.~(\ref{UkUp})
  only the average term in $U_p$ contributes to the closed major shells.}
\label{fig:master}
  \end{center}
\end{figure}

\begin{figure}[!ht]
\includegraphics[width=7 cm]{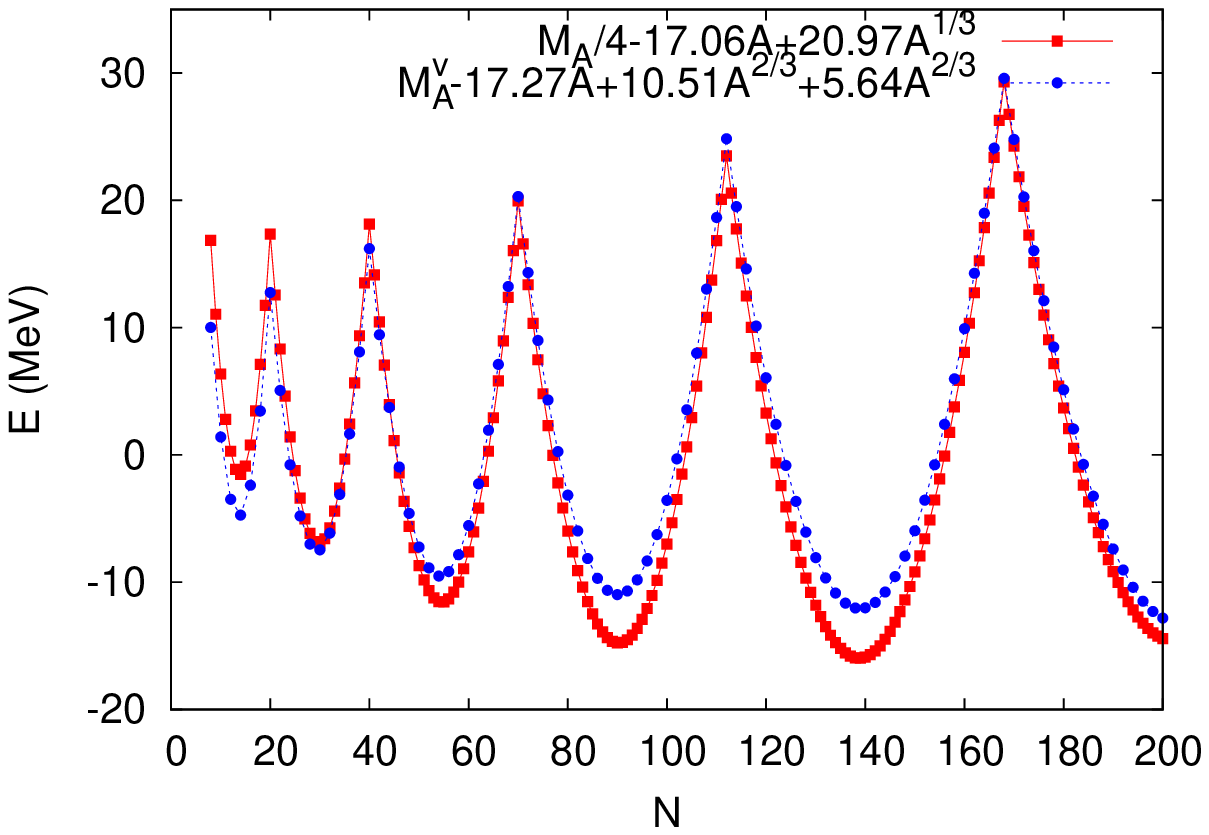} 
\includegraphics[width=7 cm]{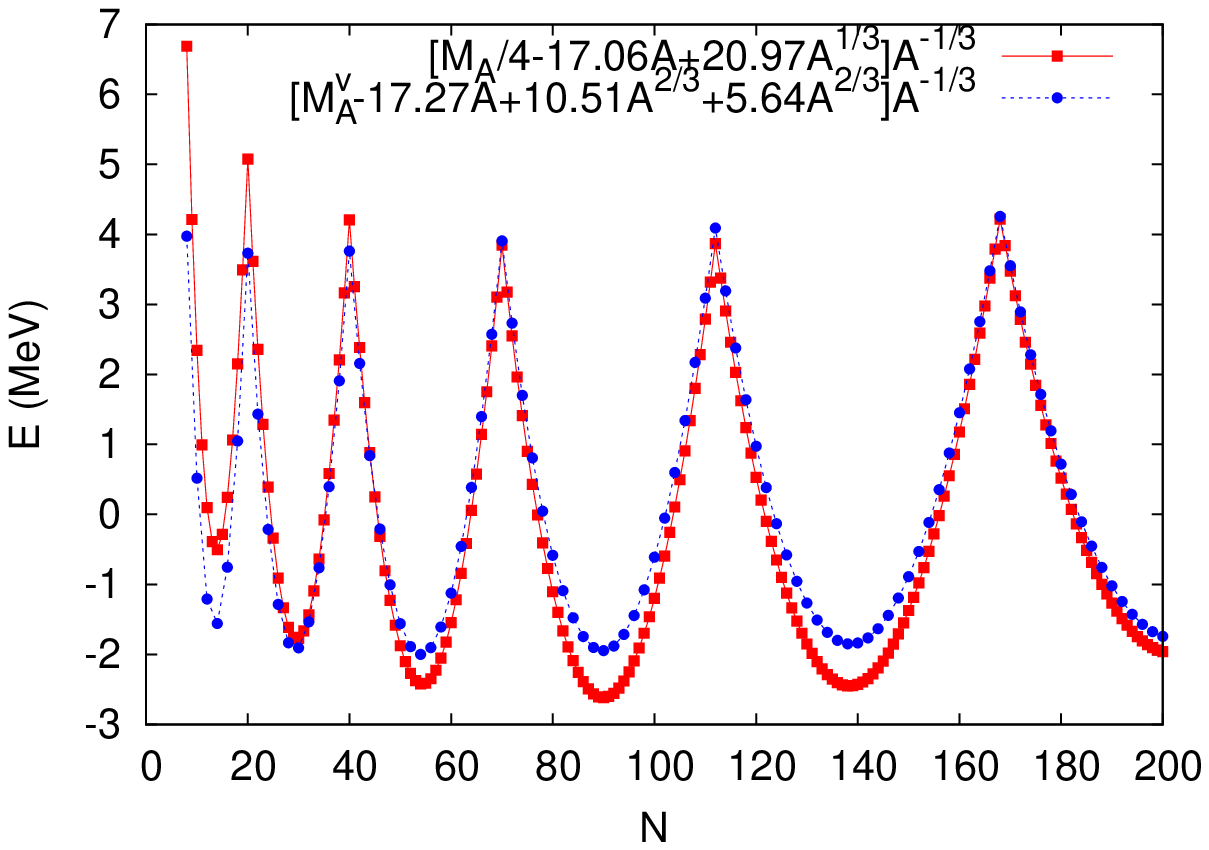} 
    \caption{Master shell effects produced by $M_A$ and $M_A^v$ for
      $t=N-Z=0$. See text.}
\label{fig:Msheff}
\end{figure}

Let us rely on the general factorization property for arbitrary
quadratic forms~\cite{Dufour}
\begin{gather}
  \sum_{x,\alpha}V_{x\alpha}\Omega_{x}\cdot \Omega_{\alpha} = \sum_{\mu}
  E_{\mu}\left
    (\sum_{k}\Omega_kU_{k_\mu}\sum_{\beta}\Omega_{\beta}U_{\beta_{\mu}}\right),
\label{factor}
\end{gather}
which we specialize to the monopole part of the Hamiltonian (detailed
discussion in~\cite{rmp05}), in which case $V$ are symmetric matrices,
diagonalized by unitary transformations $U$ and $\Omega$ stand for
isoscalar (number $m$), or isovector (isospin $T$) operators. To fix
ideas consider the result of diagonalizing the isoscalar monopole
interaction for the first 8 major oscillator shells for the chiral
N3LO interaction~\cite{n3lo} smoothed by the~$V_{{\rm low}\,k}$
procedure~\cite{vlk}. There are 36 subshells and as many
eigenvalues. One of them turns out to be strongly dominant. Within a
very good approximation its value is proportional to
$\hbar\omega\equiv 1/\rho$ and its eigenvector is independent of
it. The corresponding factor in Eq.~(\ref{factor}) can be split as
\begin{gather}
\sum_k m_kU_k= m_pU_p+\sum_k
m_k(U_k-U_p),\;U_p=\frac{\sum_kD_kU_k}{D_p},\; \sum_k D_k=\sum_k(2j_k+1)=D_p,
\label{UkUp}  
\end{gather}
where $j_k$ is the angular momentum of subshell $k$. Only the term in
$U_p$ survives when the major shell is full. It defines the collective
monopole operator in full analogy with its multipole counterparts such
as pairing and quadrupole and, again, in analogy with
them~\cite{Dufour,rmp05} we expect $U_p\propto D_P^{-1/2}$ ($\propto$
proportional to). The calculated $U_p$ are rescaled by a factor six so
as to make them of order unity and two fits are made to the $t=0$
patterns, yielding $U_p=4.25D_P^{-1/2}$, and a variant
$U^v_p=4.47D_P^{-1/2}-0.6D_p^{-1}$, which are seen to be almost
undistinguishable in Figure~\ref{fig:master}. However, the asymptotic
contributions to the master term are quite different 
$M_A\asymp 17.06A-20.97A^{1/3}$ and 
$M^v_A\asymp 17.27A-10.51A^{2/3}-5.64A^{1/3}$. They will referred to as
``asymptotics'' for short, to distinguish them from ``shell effects''.
In the first panel of Fig.~\ref{fig:Msheff} it is seen that both
approximations produce nearly the same shell effects that show as
parabolic segments bounded by HO closures at $N=$ 8, 20, 40, 70, 112
and 168. A remarkable result, in view of the very different
asymptotics. Each segment can be represented by the form
$m_p(m_p-D_p)(D_p\rho)^{-1}$, where the denominator ensures at
midshell ($m_p=D_p/2$) an amplitude $D_p/4\rho\propto
A^{1/3}$. To check that this is the correct scaling, in the second
panel of Fig.~\ref{fig:Msheff} the master shell effects of the first
panel are subject to an $A^{-1/3}$ compression.  It is seen that the
amplitudes become constant beyond $N,Z\approx 70$. The deviations from
perfect (uniform) scaling may have consequences that deserve a
comment.

Scaling is an asymptotic notion: $p+2$ and, say, $p-3$ both scale as
$A^{1/3}$ but are quite different for small $p$. They differ in a
``unit scaling'', {\em i. e.,} a factor that goes asymptotically as
unity, in our example $(p-3)/(p+2)$. The difference between the two
approximations to $U_p$ in Figure~\ref{fig:master} are of the same
type.  Fig.~\ref{fig:Msheff} provides another instance of the possible
origin of unit scalings, which in DZ work are often represented by
surface terms: typically an operator $\Gamma$ affected by
some unit scaling is replaced by $\Gamma(1+\alpha/\rho)$. More
generally one may consider rational functions {\em i. e.,} quotients
of polynomials of the same rank in $p$ (avoiding vanishing denominators,
as in Eqs.(\ref{S}) in next section).

\subsection{The HO-EI  transition}\label{sec:hoei}

In the DZ implementations 
the combination $M \rightarrow M_A+M_T$ is employed, explicitly
\begin{gather}
\label{M}
M=\frac{1}{2\rho}\left[\left(\sum_p
  \frac {m_p}{\sqrt{D_p}}\right)^2+\left(\sum_p
  \frac {t_p}{\sqrt{D_p}}\right)^2\right]  ,
\end{gather}
while the asymmetry terms are represented by $4T(T+1)/A,\;\;
(T=|t|/2)$ with the shell effects in $M_T$ simply ignored. This
neglect deserves reexamination and here we only note the
asymptotically correct form in $T(T+1)$ rather than $T^2$ is both
theoretically sound and empirically significant.

\begin{figure}[h]
\centering \includegraphics[width=9.cm]{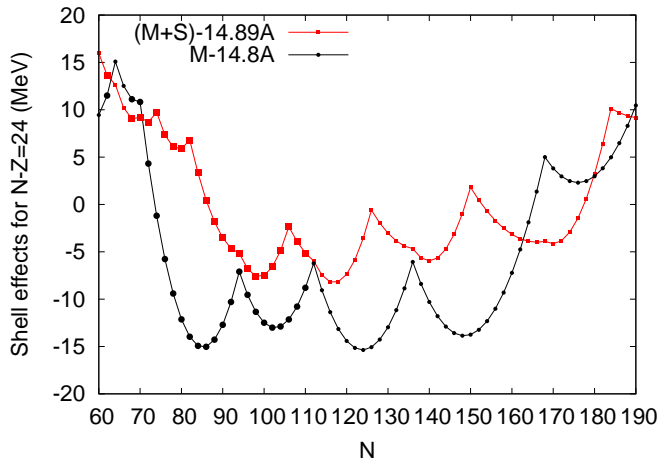}
\caption{ The evolution from HO (dots) to EI (squares) shell effects
  for $N-Z=24$ even-even nuclei.  The asymptotics are roughly
  represented by a simple $A$ term. Heavier marks refer to species
  whose masses have been measured.}
\label{fig:HO-EI}
\end{figure}

The $M$ term raises one of the outstanding problems in nuclear
physics: realistic interactions fail to produce the observed
closures~\cite{Sch06}, which contradicts the basic tenet of the
discipline that they are due to the spin-orbit force: what the famous
$l\cdot s$ term does is to make the orbit with largest angular
momentum ($j(p)$) the lowest in shell $p$. This term is indeed present
in the realistic potentials and it suggests the correct magic numbers
but it does not produce them~\cite{Sch06}. The necessary mechanism has
to be invented.
\begin{figure}[!ht]
\centering \includegraphics[width=8.cm]{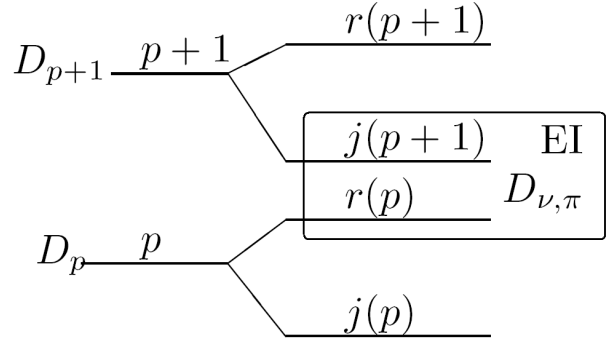}
    \caption{Harmonic oscillator and extruder-intruder
      (EI) shells. }
    \label{fig:EI}
\end{figure}
Figure~\ref{fig:EI} indicates what has to be done: To change the HO
closures (at N,Z=8,20,70 \ldots), into the observed extruder-intruder
(EI) ones at N,Z=28,50,82 and 126. As made clear by
Fig.~\ref{fig:beld} these are the only obvious ones. Therefore, the
only relevant operators must separate orbit $j(p)$ of degeneracy
$D_{j(p)}=2(p+1)$ from its partners $r(p)$ of degeneracy
$D_{r(p)}=p(p+1)$. The only one body operators that do it properly are
\beq
s_{\nu p}=\left[\frac{pn_{j_{p}}-2n_{r_{p}}}{2(p+1)}\right],~~~
s_{\pi p}=\left[\frac{pz_{j_{p}}-2z_{r_{p}}}{2(p+1)}\right],
\label{s_ops}
\eeq because they vanish at HO closures and therefore give no
asymptotic LD contribution. The $2(p+1)$ denominator is arbitrary,
$\nu$ and $\pi$ stand for neutron and proton orbits, $n$ and $z$ for
neutron and proton numbers. Duflo invented the following operator
\beqa S_{\nu}=\sum^{p_{\nu}}_{p}s_{\nu
  p}\frac{p^{2}+4p-5}{\sqrt{D_{p}}(p+2)}+ \sum^{p_{\nu}}_{p}n_ps_{\nu
  p}\frac{p^{2}-4p+5}{{D_{p}}(p+2)}, \nonumber \\
S_{\pi}=\sum^{p_{\pi}}_{p}s_{\pi
  p}\frac{p^{2}+4p-5}{\sqrt{D_{p}}(p+2)}+\sum^{p_{\pi}}_{p}z_ps_{\pi
  p}\frac{p^{2}-4p+5}{{D_{p}}(p+2)}, \nonumber \label{S} \\
S=\frac{S_{\nu}+S_{\pi}}{\rho}.~~~~~~~~~~~~~~~~~~~~~~~~~~~~~~~~~~~~~~~~~~~~~~~~
\eeqa which represents a sophisticated example of unit scaling
discussed at the end of the previous section. It leads to the
remarkable result in Fig.\ref{fig:HO-EI}, where the HO peaks are
practically erased to give way to EI ones, which, from left to right,
correspond to $N=74\, (Z=50)$, $N=82$, $N=106\; (Z=82)$, $N=126$, $N=150\;
(Z=126)$ and $N=184$. The choice of the $N-Z=24$ is arbitrary.

\subsection{The DZ10 equations. Macroscopic terms}\label{sec:dz10}

The discussion in Section~\ref{sec:orig} suggests that {\em volume}
and {\em surface} effects be treated in a single term but in DZ they
are kept separate (which turns out to be a problem, as will be
explained in the comments to Fig.~\ref{fig:DZ10M-EXP}). The terms are:
\beq
M+S ~~~{\rm and}~~~
 \frac{M}{\rho} ~~~{\rm respectively}.
\label{a1a2}
\eeq
The {\em Coulomb} term includes the charge radius $r_c$ and a surface
correction (See Ref.~\cite{DZ02} for more sophisticated treatments)
\beq
V_C=\frac{Z(Z-1)+0.76[Z(Z-1)^{2/3}]}{r_{c}};\;r_{c}=A^{1/3}
\left[1-\left(\frac{T}{A}\right)^{2}\right]
\label{coul}
\eeq
The {\em asymmetry} term uses the $T(T+1)$ form, sounder than the
usual $(N-Z)^2$
\beq
V_T=\frac{4T(T+1)}{A^{2/3}\rho},
\label{asym}
\eeq
The {\em surface asymmetry} has a small unconventional part that can
be viewed as the only---purely phenomenological---representative of
isovector shell effects,
 \beq 
V_{TS}
\equiv
\frac{4T(T+1)}{A^{2/3}\rho^{2}}-\frac{4T(T-\frac{1}{2})}{A \rho^{4}},
\label{wig}
\eeq
The {\em Pairing} term includes corrections of order $2T/A$:

\begin{tabular}{cccc}
N & Z & ~~~~~~~~~~~~~~~~~~~~~~~~~~& $V_P$ \\ \hline
even & even & & $(2 - 2T/A)/\rho $\\
even & odd & $N>Z$ &$(1 - 2T/A)/\rho$ \\
odd & even & $N>Z$ &$1 / \rho$\\
even & odd & $N<Z$ &$1 / \rho$\\
odd & even & $N<Z$ &$(1 - 2T/A)/{\rho}$\\
odd & odd & & $2T/(A\rho)$\\
\end{tabular}

Though purely phenomenological, these corrections reflect pairing
effects of Coulomb and nuclear isospin breaking
origin~\cite{Zuk02}. They ensure that the four mass sheets are
described with equal accuracy. The $1/\rho$ scaling in $A^{-1/3}$
replaces the now deprecated $A^{-1/2}$.

The  contribution to the binding energy is thus defined by six terms,
of which the last four are very close to the usual LD form:

\beqa \left\langle H_{m}\right\rangle = a_{1} \,
(M+S) -a_{2}\, \frac{M}{\rho} -
a_{3} \, V_C - a_{4} \, V_T + a_{5} \, 
V_{TS}
+ a_{6} \, V_P .
\label{dz_macro}
\eeqa
\subsection{Microscopic sector. Anomalous spherical terms}\label{sec:ano1}
The EI spaces defined by the macroscopic (macro) sector are treated as
model spaces in which to perform Shell Model calculations.  The issue
is discussed in Ref.~\cite{Zuk94}. Here we offer a compact summary.

To estimate the energies of an exact wavefunction
$|\overline{0}\rangle$ we write it as an unperturbed part $|0\rangle$
acted upon by $k$-body scalar correlation operators $\hat{A_k}$ (the
evolution operator $\exp{(iHt)}$ is a scalar). The resulting energy has
the form given in Eq.(\ref{eq:dress1}), where we separate a diagonal
part given by the monopole Hamiltonian $H_m$ and a correlation one
involving the multipole Hamiltonian $H_M$~\cite{Dufour,rmp05}. Both
$H_m$ and $H_M$ are effective operators adapted for work in the
valence spaces. Note that $\hat{A_1}$ cannot contribute since the
space does not allow scalar particle-hole excitations (no two orbits
have the same $j$ and parity).
\begin{gather}
  \label{eq:dress1}
  |\overline{0}\rangle=(1+\sum_k \hat{A_k})|0\rangle\Longrightarrow
  E=\langle 0|H_m|0\rangle+\langle 0|H_M\hat{A_2}|0\rangle
\Longrightarrow\\   \label{eq:dress2}{\rm terms\; of\; type:\; macro;}\quad
  \frac{m_v\bar{m_v}}{D_v\rho};\quad
  \frac{m_v\bar{m_v}(m_v-\bar{m_v})}{D_v^2\rho}; \quad
  \frac{m_v^{(2)}\bar{m_v}^{(2)}}{D_v^3\rho}
\end{gather}
In Eq.(\ref{eq:dress2}) we use the notations $m^{(2)}_v=m_v(m_v-1)$
and $\bar{m_v}=D_v-m_v$ to write the form of the possible
contributions in a valence space $v$ of degeneracy $D_v$.  The macro
term is solely responsible for the closed shells at
$m_v=\bar{m_v}=0$. The origin of the quadratic term can be both
monopole (as in Fig.~\ref{fig:Msheff}) and multipole (e. g. the
pairing interaction in its simplest form). We have no argument to
ascribe the cubic terms to configuration mixing.  At the time the DZ
model(s) were formulated the possibility that it could be due to a
genuine monopole three body force was thought to be unlikely; nowadays
it has become a certainty that such forces are essential. In
section~\ref{sec:macro} we shall find clues indicating that the
quadratic term owes much to configuration mixing and that the cubic
one must be due to a genuine three body force whose determinant role
in the HO-EI transition will be established in
Section~\ref{sec:3b}. We are left with the quartic correlation term
whose form is dictated by the remark that $H_M\hat{A_2}$ is a four
body operator that must vanish at $m_v=\bar{m_v}=0$ or 1 because only
$H_m$ acts at these points.

In Eq.~(\ref{eq:dress2}) the denominators are chosen so as to ensure
correct $A^{1/3}$ scalings. Unfortunately, an original error
(explained in the discussion of Fig.~\ref{fig:DZ10M-EXP}) leads to the
need of anomalous scalings for the ``spherical'' terms that operate in
the EI valence spaces. Using $\nu$ and $\pi$ indeces for neutron and
proton quantities, they take the form
\begin{gather}\label{eq:s3}
  s_3=\frac{1}{\rho}\left[\frac{n_{\nu}\bar{n}_{\nu}(n_{\nu}-\bar{n}_{\nu})}{D_{\nu}}
+\frac{n_{\pi}\bar{n}_{\pi}(n_{\pi}-\bar{n}_{\pi})}{D_{\pi}}\right],
  \\ \label{eq:s4} s_4=\frac{1}{\rho}\left[2^{(\sqrt{p_{\pi}}+\sqrt{p_{\nu}})} \cdot
    \left(\frac{n_{\nu}\bar{n}_{\nu}}{D_{\nu}}\right) \cdot
    \left(\frac{n_{\pi}\bar{n}_{\pi}}{D_{\pi}}\right)\right].
\end{gather}

leading to contributions to the binding energy
\beq \left\langle H_{s}\right\rangle=a_{7} \, s_{3} -
a_{8} \, \frac{s_{3}}{\rho} +a_{9} \, s_{4}.
\label{dz_sph}
\eeq

\subsection{Deformation term}\label{sec:def1}
That the onset of deformation is due to the promotion of four neutrons
and four protons from an $r$ shell to the next major HO shell is
something that can be read from Nilsson diagrams as pointed out in
\cite{Zuk94}. The loss of macroscopic (monopole) energy is upset by
the gain due to the quadrupole force, simulated by a specific quartic
operator (which scales correctly). Calling generically $n'=n-4,
~\bar{n'}= \bar{n}+4$, its form and contribution to the energy are

\beq
d_4=\frac{1}{\rho}\left[\left(\frac{n'_{\nu}\bar{n'}_{\nu}}{D^{3/2}_{\nu}}\right) \cdot
\left(\frac{n'_{\pi}\bar{n'}_{\pi}}{D^{3/2}_{\pi}}\right)\right], ~~~
\left\langle H_{d}\right\rangle= a_{10} \, d_4.
\label{dz_def}
\eeq

The idea was very successfully incorporated in the original DZ fits,
and is at the origin of the spherical description of rotational
nuclei~\cite{ZRPC95,rmp05,Zuk08}, by now firmly established, which
goes well beyond Eq.~(\ref{dz_def}) and points to its limitations. The
valence space involves two contiguous major HO shells in which good
approximations close to Elliott's SU3 symmetry operate.
Eq.~(\ref{dz_def}) simulates well the mechanism at the beginning of
the deformed regions by staying in the EI space where the $r$
particles represent the lower shell and the $j$ orbit the upper
one. It fails as soon as the $r$ orbits are
full. Fig.~\ref{fig:t32-evol} in Section~\ref{sec:ano2} will provide
an example of the problem.

\section{The fits}\label{sec:fits}
Two calculations are made for each nucleus. Both include the
macroscopic contribution plus either the anomalous spherical terms or
the deformed one.

The $a_{i}$ coefficients are varied to minimize the root mean square
deviation (RMSD) between the predicted binding energies $BE_{\rm
  th}(N,Z)$ and the experimental ones $BE_{\rm exp}(N,Z)$, reported
in~\cite{AME03}, modified so as to include more realistically the
electron binding energies as explained in Appendix A of Lunney,
Pearson and Thibault~\cite{Lunn03}.  For each nucleus a spherical and
a deformed calculation are made, and the one with largest binding is
selected:

 \beqa BE_{th}=\left\langle H_{m}\right\rangle + \left\langle
H_{s}\right\rangle \hbox{~~~~if Z $<$ 50} \nonumber \\
BE_{th}=\left\langle H_{m}\right\rangle+max \left(\left\langle
H_{s}\right\rangle,\left\langle H_{d}\right\rangle\right) \hbox{~~~~if
  Z $\geq$ 50}
\label{dz10}
\eeqa

\begin{equation}
{\rm RMSD}=\left\{\frac{{\sum\left[BE_{\rm exp}(N,Z)-BE_{\rm
        th}(N,Z)\right]^2}}{N_{nucl}}\right\}^{1/2}.
\end{equation}
$N_{nucl}$ is the number of nuclei for $N,Z\geq8$. The minimization
procedure uses the routine Minuit \cite{Minuit}.

\subsection{Macroscopic sector}~\label{sec:macrofit}
\begin{figure}[ht]
\centering\includegraphics[width=9 cm]{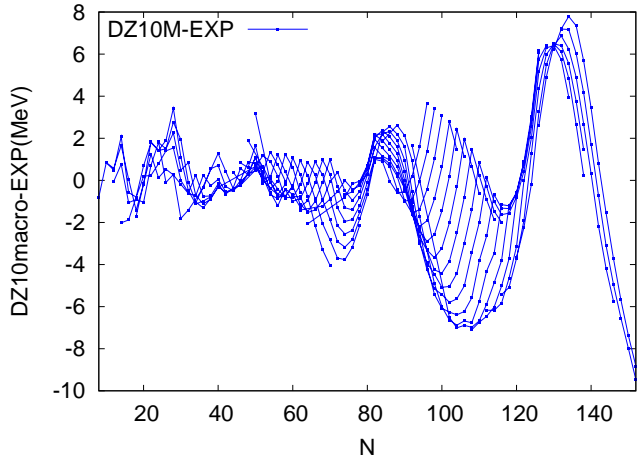} 
    \caption{Differences between the binding
      energies predicted by DZ10 macro,
      Eq.~(\ref{dz_macro}) and the experimental ones. Even-even
      nuclei (RMSD=2.86 MeV). Lines join points at constant $t=N-Z$}
\label{fig:DZ10M-EXP}
\end{figure}

The heavy task of the macroscopic sector is to ensure asymptotically a
LD form and at the same time generate shell effects that move the HO
closures to EI ones. As the Coulomb, asymmetry and pairing terms are
already of LD type---within minor provisos---the task of generating
shell effects falls on the first two terms in Eq.~(\ref{dz_macro}) in
charge of bulk volume and surface contributions, as we have seen in
Fig.~\ref{fig:HO-EI}. Table~\ref{ajuste} follows the evolution of the fits as
the coefficients are turned on.
\begin{table} [!ht]
\centering
\caption{The macroscopic coefficients, with their associated mean and
  RMS errors.}
\label{ajuste}
	\begin{tabular}{|c|c|c|c|c|c|c|c|}
\hline {\bf Op}&$a_i$ & \multicolumn{ 6}{|c|} {}\\ \hline
$ M+S$   & $a_{1}$ & 9.3980 &  5.2535  & 4.6971   & 16.6714 & 17.3653 & 17.5337\\

$M/\rho$ & $a_{2}$ & 0.0000 & -22.9448 & -24.6563 & 11.8321 & 14.7737 & 15.4380\\

$V_C$    & $a_{3}$ & 0.0000 & 0.0000   & -0.0405  & 0.6680  & 0.6870  &  0.6946\\

$V_T$    & $a_{4}$ & 0.0000 & 0.0000   & 0.0000   & 26.1232 & 35.6940 & 36.1628\\

$V_{TS}$ &$a_{5}$  & 0.0000 & 0.0000   & 0.0000   & 0.0000  & 47.0442 &  48.4240\\

$V_P$    &$a_{6}$  & 0.0000 & 0.0000   & 0.0000   & 0.0000  & 0.0000  &  5.0943\\
\hline 
\multicolumn{2}{|c|}{\bf RMSD} &64.062 & 29.190 & 29.063 & 4.232 & 2.934 & 2.852\\
\multicolumn{2}{|c|}{\bf mean} & 21.032 & -2.222 & -2.359 & -0.163 & -0.043 & -0.053\\ 
\hline
\end{tabular}  
\end{table} 
If we judge by these numbers, DZ10 does hardly better than any
standard LD fit. However, this is not an LD fit and surprises are in
store. Fig.~\ref{fig:DZ10M-EXP} displays the differences
between the experimental and theoretical binding energies, calculated
using only the macroscopic terms, Eq.~(\ref{dz_macro}). The pattern is
both disconcerting and reassuring. It blows up at around $N\approx 80$
and there is no trace of the expected $A^{1/3}$ scaling. Things go
much faster and the same time become nicely systematic. The scaling
problem (the ``original error'' mentioned in Section~\ref{sec:ano1})
can be traced to Section~\ref{sec:orig} where it was shown that
different variants of the master term have the same shell effects but
different asymptotics; the leading $M+S$ term in Eq.~(\ref{a1a2}) has
the correct asymptotics in $A$, no $A^{2/3}$ surface term and a
$A^{1/3}$ part whose effects are sizeable (about 100 MeV around
$A=100$). The $M/\rho$ surface contribution in
Eq.~(\ref{a1a2})---which also contains HO shell effects---is left in
charge of correcting the $A^{1/3}$ ``drift''. An artificial procedure
that explains the need of anomalous scalings but does not hide the
systematic behavior they have to simulate.

\subsection{The microscopic sector}\label{sec:micro}
In Table~\ref{ajuste2} the full set of coefficients $a_i$, with their
associated mean and RMS errors, is presented. All fits include the
six macroscopic terms, plus some, or all, the microscopic ones, $a_7$
to $a_{10}$. It is clear that the three spherical terms can provide
a reasonable good fit, with an RMSD of 0.72 MeV, but the three terms
must be present, acting together. With only two of them the RMS error
is larger than 2 MeV, as is the case if only the term associated with
``deformed'' nuclei is employed.  When the 10 terms are active Eq.~(\ref{dz10}) 
fits the AME03 set with an RMSD of 0.55 MeV.

\begin{table} [!htb]
\centering
\caption{The full set of coefficients, with their associated mean and
  RMS errors.}
\label{ajuste2}
\begin{tabular}{|c|r|r|r|r|r|c|c|}
\hline {\bf Op}&$a_i$ & \multicolumn{ 6}{|c|}{} \\ \hline 
$ M+S$    & $a_{1}$ & 17.492 & 17.542 & 17.769 & 17.778 & 17.770 & 17.766\\
$M/\rho$  & $a_{2}$ & 15.284 & 15.507 & 16.258 & 16.355 & 16.210 & 16.314\\
$V_C$     & $a_{3}$ & 0.693  & 0.694  & 0.708  & 0.708  & 0.707  & 0.707\\        
$V_T$     & $a_{4}$ & 35.513 & 35.721 & 38.354 & 37.480 & 38.080 & 37.515\\
$V_{TS}$  & $a_{5}$ & 45.836 & 46.653 & 56.734 & 53.232 & 55.394 & 53.351\\
$V_P$     & $a_{6}$ & 5.414  & 5.275  & 5.5361 & 6.373  & 5.269  & 6.199\\
$s_3$     & $a_{7}$ & 0.062  & 0.448  & 0.000  & 0.390  & 0.000  & 0.478\\
$s_3/\rho$& $a_{8}$ & 0.000  & 2.106  & 0.000  & 1.763  & 0.000  & 2.183\\
$s_4$     & $a_{9}$ & 0.000  & 0.000  & 0.0215 & 0.025  & 0.000  & 0.022\\
$d_4$     & $a_{10}$& 0.000  & 0.000  & 0.000  & 0.000  & 37.568 & 41.338\\
\hline 
\multicolumn{2}{|c|}{\bf RMSD} & 2.443 & 2.293 & 2.028 & 0.717 & 2.280 & 0.554\\
\multicolumn{2}{|c|}{\bf mean} & -0.040 & -0.043 & -0.035 & 0.002 &-0.031 & 0.000\\
 \hline
\end{tabular}  
\end{table}
\begin{figure}[ht]
\begin{minipage}{.49\linewidth}
\centering \includegraphics[width=7cm]{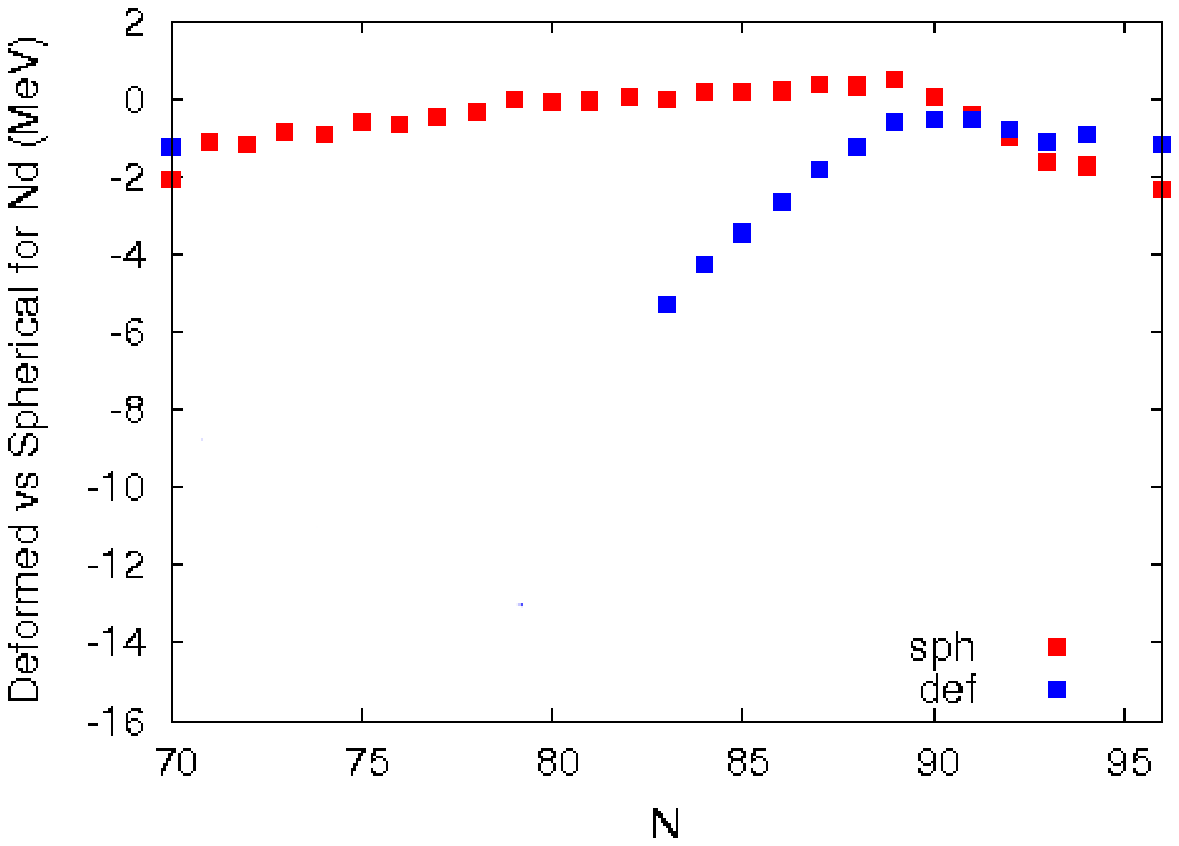} \center{a}
\end{minipage}
\begin{minipage}{.49\linewidth}
\centering \includegraphics[width=7cm]{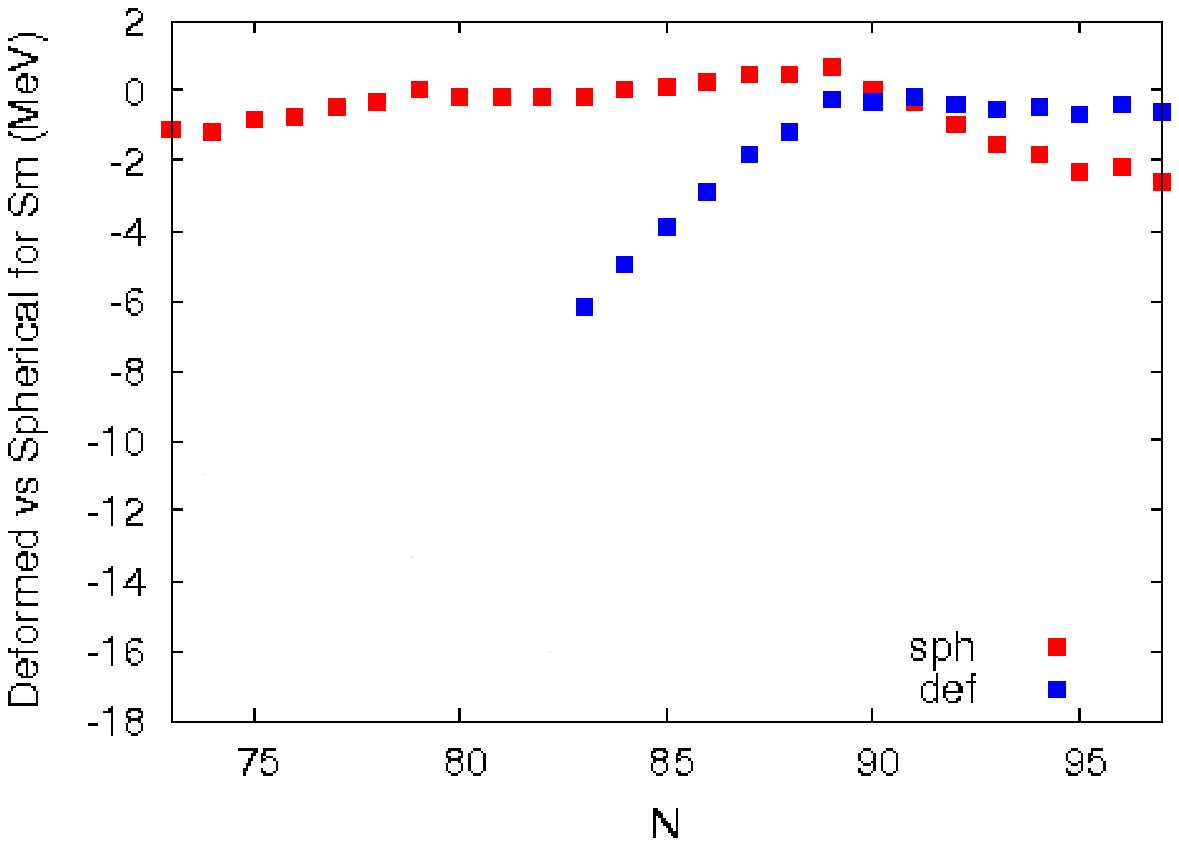} \center{b}
\end{minipage}
\begin{minipage}{.49\linewidth}
\centering \includegraphics[width=7cm]{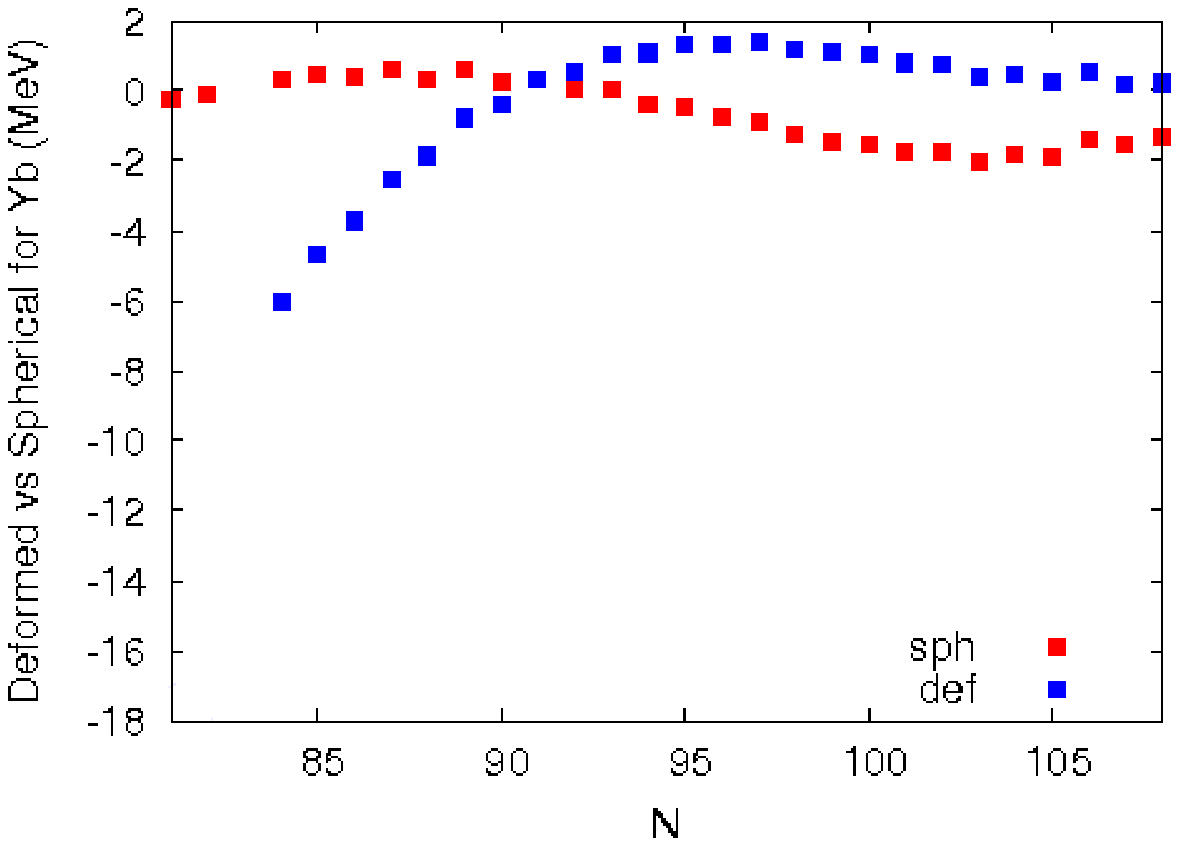} \center{c}
\end{minipage}
\begin{minipage}{.49\linewidth}
\centering \includegraphics[width=7cm]{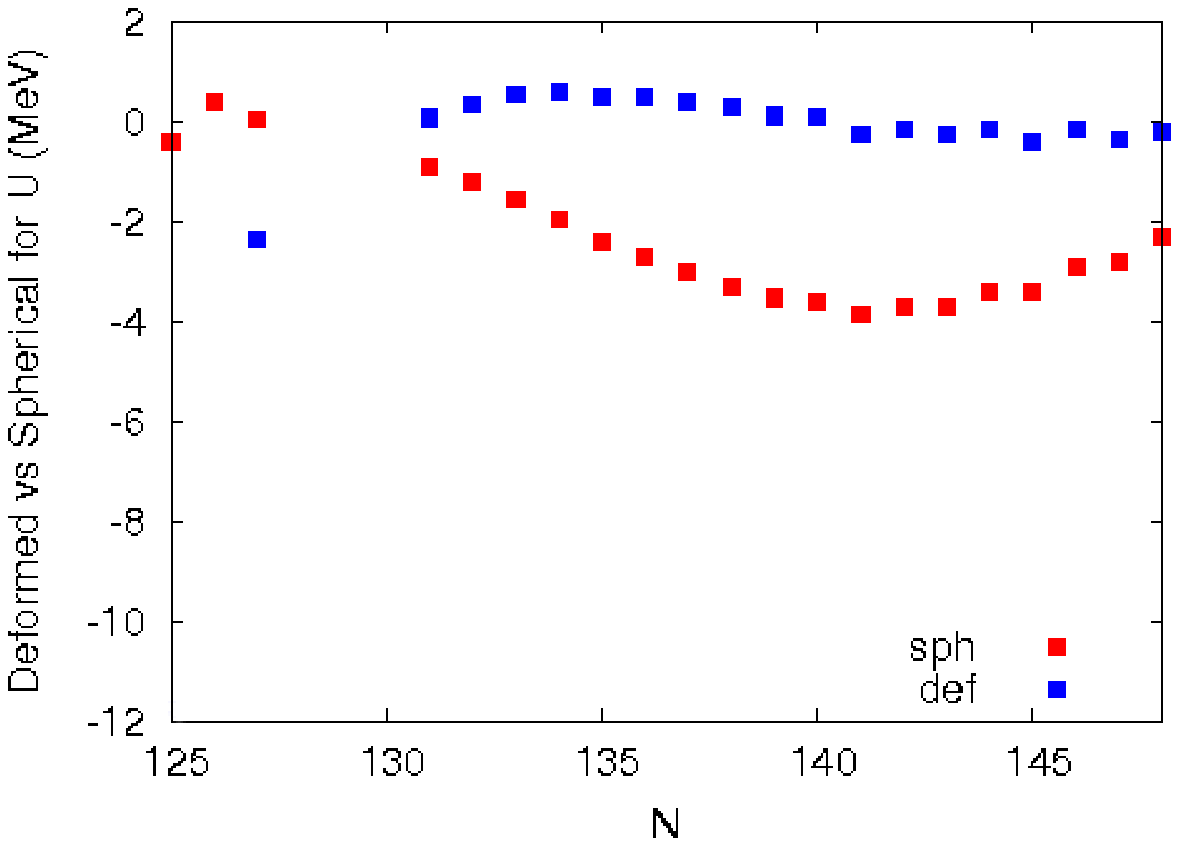}
\center{d}
\end{minipage}
\caption{Transition from spherical to deformed for the following chains of isotopes (a) Neodymium ($_{60}$Nd); (b) Samarium ($_{62}$Sm); (c) Ytterbium ($_{70}$Yb) and (d) Uranium ($_{92}$U).}
\label{fig:isotopes}
\end{figure}
From the the discussion of Fig.~\ref{fig:DZ10M-EXP} we understand both
why anomalous scalings become necessary and why the terms of
Eq.~(\ref{dz_sph}) manage to be so efficient. The factor
$2^{\sqrt{p_{\nu}}+\sqrt{p_{\pi}}}$ in Eq.~(\ref{eq:s4}), though
atrocious, is a symptom of the difficulty of accommodating
Fig.~\ref{fig:DZ10M-EXP} with a single scaling. Eliminating it costs
only some 50 keV. On the other hand, forcing the $A^{1/3}$ scaling
leads to losses of some 200 keV. We shall examine closely the action
of these terms in Section~\ref{sec:ano2}.

\subsection{The role of deformation}\label{sec:def2}
From the 2149 nuclei with masses reported in AME03 and N,Z $\ge$ 8, 1827
are found to be spherical and 322 deformed.
Figure \ref{fig:isotopes} shows the DZ10 deformed and
spherical binding energies subtracted from the experimental ones for
four chains of isotopes. The crossings signal the onset of
deformation, which reproduces perfectly the N=90 transition region.
Note as a pleasant result the correct inclusion of $^{130}$Nd among
deformed nuclei. It should be noted though that DZ10 underestimates
the number of nuclei involved, as anticipated in
Section~\ref{sec:def1} and illustrated in Section~\ref{sec:ano2}.

\section{Predictive power and stability of DZ10}\label{sec:predictive}
\begin{figure}[!ht]
		\dib{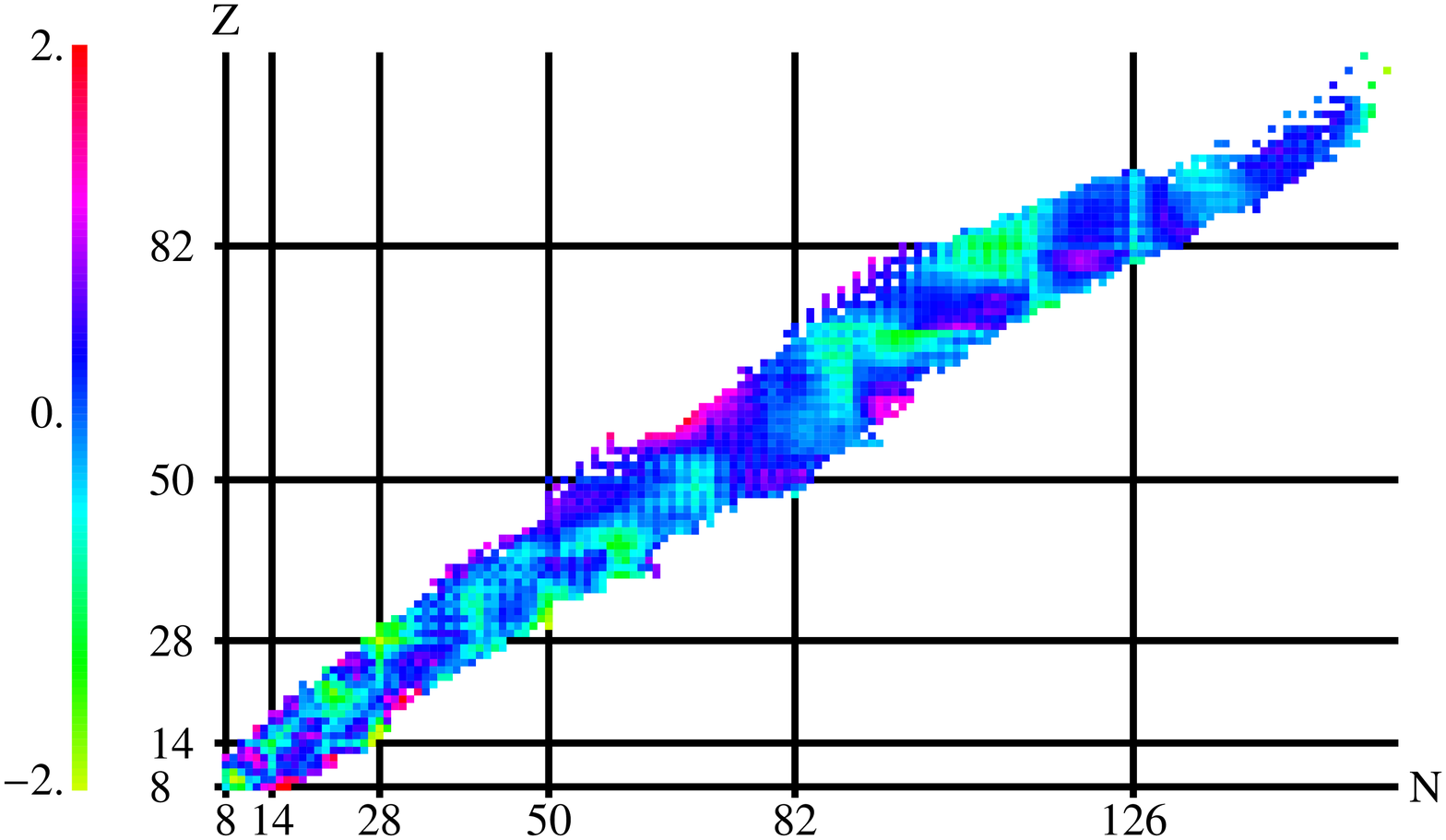}
                \dib{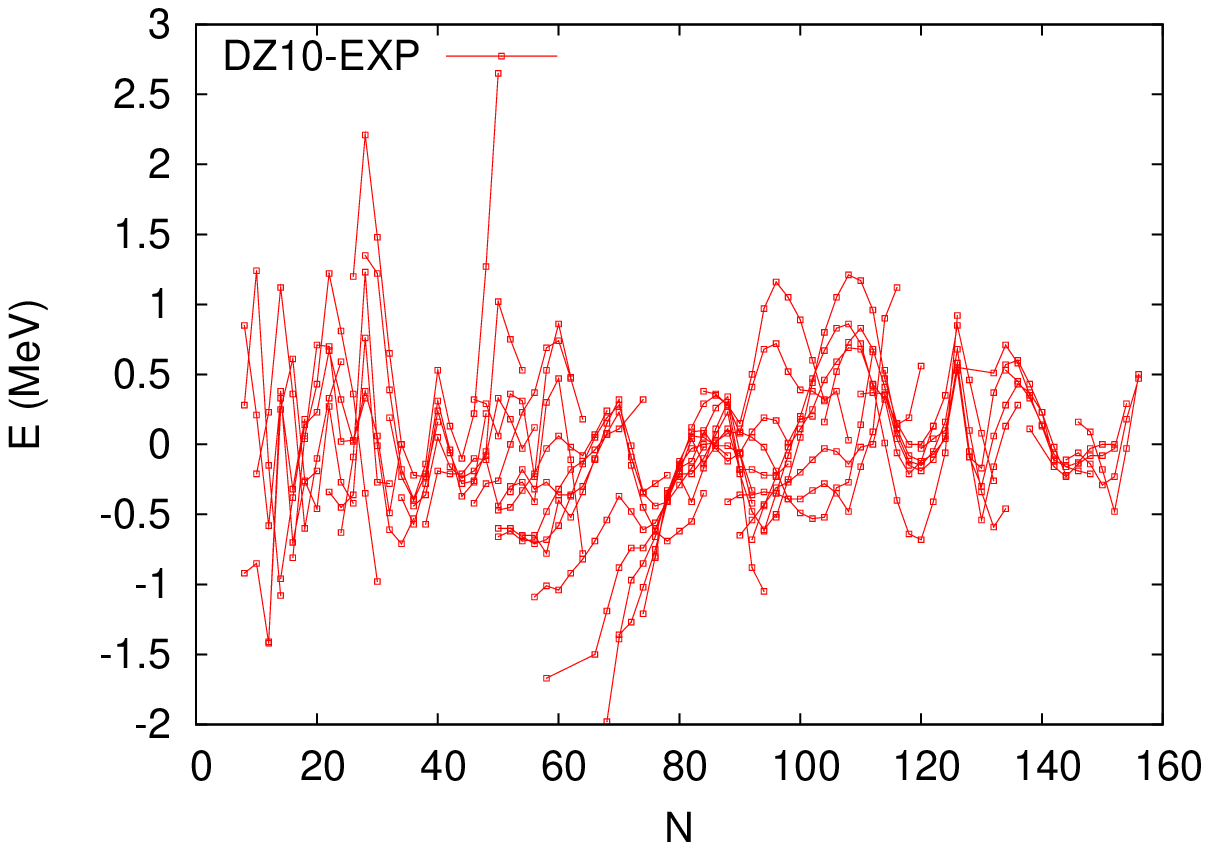}
    \caption{Differences between the experimental binding energies and
      those calculated with the full fit. (RMSD= 0.55 MeV)}
\label{fig:macro+micro}
\end{figure}

The quality of a fit depends not only on its RMSD but also on
its error pattern: The closer the latter is to a uniform random
distribution the better. As we have seen in Fig.~\ref{fig:DZ10M-EXP},
systematic deviations are not necessarily a bad thing. Strong isolated
discrepancies are more serious.  Figs.~\ref{fig:macro+micro} shows
that both effects are present in the full fit. To assess their impact
we propose to follow some recent work~\cite{Men08} where a number of
tests were introduced which
probe the ability of nuclear mass models to extrapolate, all based on
nuclear masses taken from ref.~\cite{AME03}.  The tests are performed
for the 2149 nuclei with $N\geq8,Z\geq8$ or the 1825 nuclei with
$N\geq28,Z\geq28$.  They are:
\begin{itemize}
\item{\bf AME95-03}: The subset of nuclei with measured masses in the
  AME95 compilation~\cite{AME95} is fitted.  This test was used in
  ref.~\cite{Lunn03} to compare predictions of different models.  In
  this work the actual masses used in the fit were taken from AME03,
  only the set of nuclei to be fitted is based on AME95.
\item{\bf Border region}: Nuclei which are furthest removed from
  stability are excluded from the fit and subsequently predicted by
  extrapolation.
\item{\bf Lead region}: Nuclei with mass number $A\leq 160$,
  170, 180, 190 or 200 are fitted and the remaining ones, which always
  include the region around $^{208}$Pb, are predicted by
  extrapolation.
\end{itemize}
\begin{table}[!htbp]
\centering
\caption{RMSD of the different test applied to DZ10.}
\label{test}
\begin{tabular}{|c|c|c|c|c|}
\hline Test & \multicolumn{ 2}{|c}{ $N,Z\geq8$ } & \multicolumn{
  2}{|c|}{ $N,Z\geq28$ } \\ \hline & fit & prediction & fit &
prediction \\ \hline Full set & 0.5537 & & 0.4819 & \\

  AME95-03 & 0.5076 & 0.7823 & 0.4469 & 0.6276 \\

     Border & 0.4988 & 0.8212 & 0.4540 & 0.6527 \\

      $A\leq160$ & 0.5787 & 0.9049 & 0.4823 & 1.2330 \\

      $A\leq170$ & 0.5847 & 0.7740 & 0.4966 & 1.2889 \\

      $A\leq180$ & 0.5708 & 0.7711 & 0.4859 & 1.0481 \\

      $A\leq190$ & 0.5855 & 0.5611 & 0.5098 & 0.5010 \\

      $A\leq200$ & 0.5844 & 0.5014 & 0.5054 & 0.4728 \\ \hline
\end{tabular}  
\end{table}

The number of nuclei with predicted masses ranges from 371 in AME95-03
to 810 for $A\leq 160$.  The RMSD of the 16 fits are summarized in
Table~\ref{test}. In the first line the RMSD of the fits of the whole
set of nuclei are presented. The fit is better when the lighter nuclei
are excluded. Notice that, while the RMSD of the fits is slightly
smaller for the subsets AME95 and border, as compared with the full
set of nuclear masses, when the region around $^{208}$Pb is excluded
from the fit the RMSD are always larger. Also for the three subsets $A
\leq160,~ 170,~ 180$ the RMSD of the predictions is noticeably larger
than the RMSD of the fits. The situation changes drastically for $A
\leq 190,~ 200$, where the RMSD of the predictions are slightly smaller
than those of the fits.

\begin{table}[!htbp]
\centering
\caption{Coefficients of the DZ10 mass formula obtained from the best
  fit of the full data set, their average, dispersion and fractional
  variation from the 16 fits.}
\label{stat}
\begin{tabular}{|c|ccccc|}
\hline Coefficient & $a_1$ & $a_2$ & $a_3$ & $a_4$ & $a_5$ \\ \hline
full set & 17.766 & 16.313 & 0.707 & 37.514 & 53.344 \\

      mean & 17.773 & 16.332 & 0.707 & 37.327 & 52.277 \\

   $\sigma$ & 0.011 & 0.027 & 0.001 & 0.333 & 1.759 \\

      $100\,\sigma/mean$ & 0.06 & 0.17 &
    0.19 & 0.89 & 3.37 \\ \hline Coefficient & $a_6$ & $a_7$ &
    $a_8$ & $a_9$ & $a_{10}$ \\ \hline full set & 6.1985 & 0.4784 &
    2.1831 & 0.0216 & 41.3423 \\

      mean & 6.2206 & 0.4853 & 2.1992 & 0.0218 & 40.6985 \\

    $\sigma$  & 0.1044 & 0.0426 & 0.2045 & 0.0006 & 1.4593 \\

     $100\,\sigma/mean$  & 1.68 & 8.77 &
    9.30 & 2.95 & 3.59 \\ \hline
\end{tabular}  
\end{table}
Associated with the 16 fits described above are 16 sets of coefficients
$\{a_i\}$. They allow for a statistical analysis of their mean value,
their root mean square deviation, and the percentage of variation,
estimated as $100\,\sigma/\left|mean\right|$. These numbers are
reported in Table~\ref{stat}. In the first line the values obtained
for the full fit of masses wit $N,Z \geq 8$ are included for
comparison.  The six coefficients employed in the macroscopic terms
vary less than 3\%, the microscopic coefficients are found to vary up
to 10\%. In any case, it is clear that the DZ10 is very stable, with
its coefficients varying smoothly and very moderately when the set of
data employed in the fit is changed.

\section{Analysis of the anomalous effects}\label{sec:ano2}

The anomalous spherical terms are conceptually unacceptable but
phenomenologically crucial. To gain some insight into their behavior
we choose five families of data with $t=N-Z=$8, 16, 24, 32 and 40, and
examine how the macroscopic (macro) patterns are driven to reasonable
agreement with the data (exp) in the full fit (dz10).
\begin{figure}[!ht]
\includegraphics[width=8 cm]{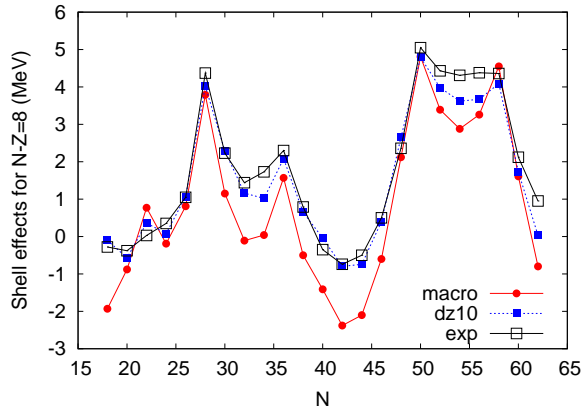} 
    \caption{Effect of anomalous terms for $N-Z$=8
 even-even nuclei referred to LD.}
\label{fig:t8-evol}
\end{figure}
Fig.~\ref{fig:t8-evol} for $t=8$ shows that macro comes close to the
exp pattern but is slightly unbound. The anomalous terms bring in the
necessary correction.
\begin{figure}[!ht]
\includegraphics[width=8 cm]{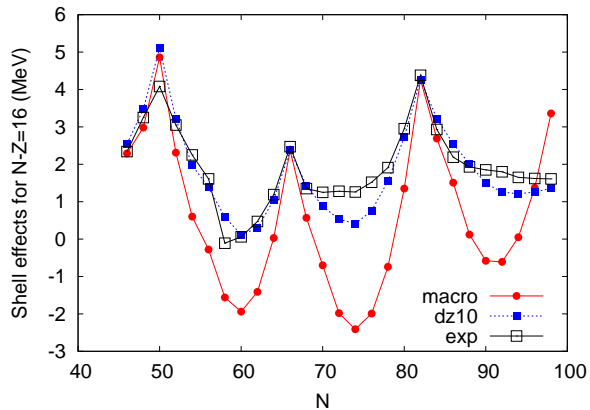} 
    \caption{Effect of anomalous terms for $N-Z$=16 even-even
     nuclei referred to LD.}
\label{fig:t16-evol}
\end{figure}

For $t=16$ in Fig.~\ref{fig:t16-evol} the situation changes
dramatically: macro is way off except at the closures. Contrary to the
$t=8$ family where the naive shell model seems valid, most of the
$t=16$ nuclei exhibit vibrational features. In principle they would
demand a separate treatment as done for the well deformed species. As
such a treatment remains to be found, DZ relies on the anomalous
terms to restore reasonable agreement.
\begin{figure}[!ht]
\includegraphics[width=8 cm]{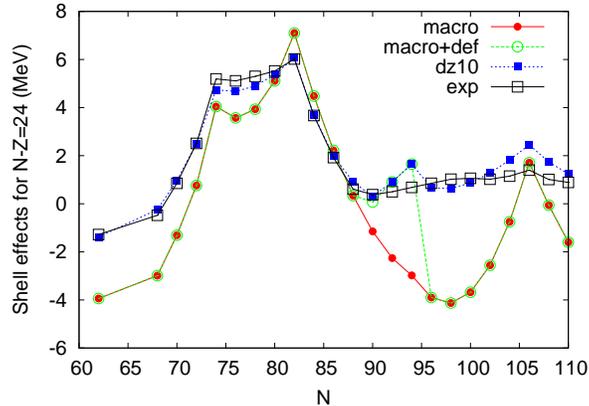} 
    \caption{Effect of anomalous terms for $N-Z$=24 even-even
     nuclei referred to LD.}
\label{fig:t24-evol}
\end{figure}

For $N-Z=24$ in Fig.~\ref{fig:t24-evol} the situation is more
complicated: macro gives a good description near closures at $N=82$
and $N=106,\, Z=82$ ($^{188}$Pb) but starts deviating strongly at $N=90$ where
deformation sets in. The inclusion of $d_4$ in Eq.~(\ref{dz_def})
(macro+def in the figure) restores agreement with data for
$N=90,92,94$ that rapidly deteriorates at $N=96$. The inclusion of the
anomalous terms brings back reasonable agreement. This provides
a clue that supplements what was found for $t=16$.

DZ does an excellent job at describing the onset of deformation at
$N=90$ ($^{156}$Dy) through 4n-4p jumps, an idea vindicated by later
work~\cite{ZRPC95,rmp05,Zuk08}.  However, deformed regions not only
start somewhere, they also end somewhere, in our case at the
$^{188}$Pb weak closure. In between, low lying $\gamma$ bands indicate
the need to go beyond 4n-4p jumps. DZ does not include this option and
leaves the anomalous ``spherical'' terms in charge of deformation
after $d_4$ fails to do it. Note that contrary to the $t=16$
``vibrators'' the necessary tools to go beyond 4n-4p jumps are
available, as explained in Section~\ref{sec:def1}.
\begin{figure}[!ht]
\includegraphics[width=8 cm]{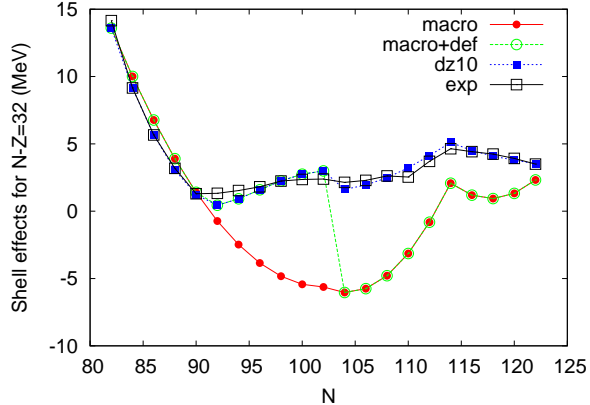} 
    \caption{Effect of anomalous terms for $N-Z$=32 even-even
     nuclei referred to LD.}
\label{fig:t32-evol}
\end{figure}

Something similar occurs for $N-Z=32$ in Fig.~\ref{fig:t32-evol}:
macro is good at first, then deformation sets in and is well described
by def for a while but then the anomalous terms take over in a region
that remains well deformed. They even correct nicely what is missed by
macro at the $N=114,\, Z=82$ closure.
\begin{figure}[!ht]
\includegraphics[width=8 cm]{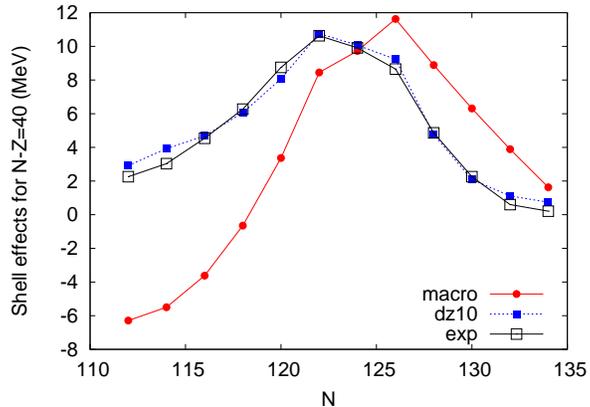} 
    \caption{Effect of anomalous terms for $N-Z$=40 even-even
     nuclei referred to LD.}
\label{fig:t40-evol}
\end{figure}
The situation becomes baffling for $t=40$ in Fig.~\ref{fig:t40-evol}
because of the uncanny capacity of the anomalous terms to restore
agreement with data that appeared compromised by the distorted
macroscopic pattern.

The anomalous terms can deal with practically everything, including
deformation. This is achieved by borrowing the sound idea of simple
polynomial forms and subverting it by introducing unacceptable
scalings. Among the many things they do, the anomalous terms correct
shortcomings of the macroscopic contribution, as seen in the $Z=82$
closure for $t=32$ and in the $t=40$ case. We have already traced the
scaling flaws of the macroscopic terms in the discussion of
Fig.~\ref{fig:DZ10M-EXP} but its worth pushing their analysis in a
totally different direction.

\section{Analysis of the macroscopic term. GEMO}\label{sec:macro}

Rather than attempting to correct directly the scaling flaws of DZ10,
we shall compare the model with a complementary approach in which it
is very simple to force correct asymptotics, and has the enormous
advantage of being practically parameter free.

Ideally, the macroscopic-microscopic separation should mirror the
monopole-multipole separation of the realistic
interactions~\cite{Zuk94,Dufour,rmp05}. We recall that the monopole
part provides the natural definition of the unperturbed Hamiltonian as
it contains all the number and isospin operators and all that is
needed for a spherical Hartree Fock variation. Our purpose is to
examine to what extent the macroscopic DZ10 terms are consistent with
a monopole Hamiltonian derived without any reference to masses;
defined in~\cite{DZ99} (DZII or GEMO (GEneral MOnopole) after the name
of the code~\cite[under item containing~\cite{DZ99}]{dzcode}) by
fitting (quite well) all single particle and single hole states on
doubly magic nuclei (the $cs\pm 1$ set). Some details that supplement
the original paper can be found in~\cite{Zuk08,rmp05}.
\begin{figure}[b]
\centering\includegraphics[width=9 cm]{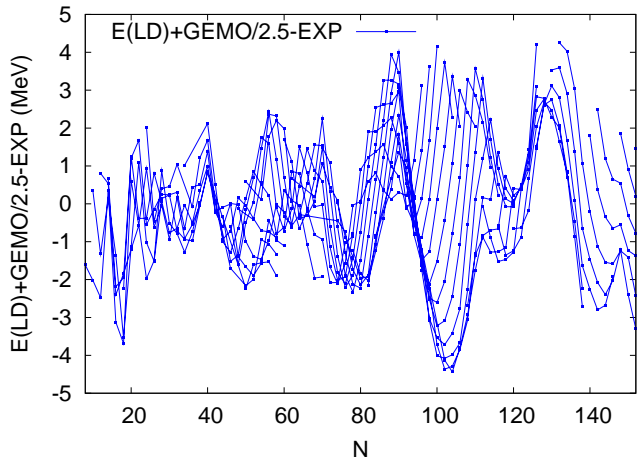} 
\caption{Comparison between GEMO and experimental masses for even-even
  nuclei. 
RMSD=1.69  MeV.  Lines join points at constant $t=N-Z$. See
  text}
\label{fig:GEMO-EXP}
\end{figure}
GEMO is fully equivalent to a mean field calculation of monopole shell
effects. It decouples approximately the macroscopic part by replacing
the master term by the combination $M_A-4K$ which has no $A$ and
$A^{2/3}$ contributions and produces shell effects similar to those of
Fig.~\ref{fig:Msheff}. $M_A$ is from Eq.~(\ref{master}) and $K$ is the
kinetic energy affected by an $\hbar\omega_0\approx 11$ denominator
chosen to reproduce the correct coefficient for the  $M_A-4K$
term.
\begin{gather}
 K=\frac{\hbar \omega}{2\hbar \omega_0}\sum_p m_p(p+3/2)
\Longrightarrow
\frac{\hbar \omega}{4\hbar \omega_0}(p_f+3)^{(3)}(p_f+2)
\label{K}
\end{gather}
Asymptotically we obtain from a fit $(M_A-4K)\asymp -28.04
A^{1/3}-9.15t^2/A+9.43$; and from now on GEMO stands for the properly
subtracted value. To compare with experiment {\em we subject GEMO to a
  2.5 contraction} (the only free parameter, to be examined later) and
we add to it $E(LD)$ from Eq.(\ref{ld}). In the calculation the
filling order of the orbits has been kept fixed throughout, according
to an $l\cdot s$ law with an $l\cdot l$ modulation. Much better
results are expected if the proper procedure is followed by
determining variationally the filling order for each nucleus. In spite
of this shortcoming the results shown in Fig.~\ref{fig:GEMO-EXP} have
substantially smaller RMSD than those for DZ10 in
Fig.~\ref{fig:DZ10M-EXP} 
(1.69 {\em vs} 2.86 MeV. 
Beware of y-scale differences between the
figures). However, for $N\lessapprox 80$ both models have very much the
same RMSD$\approx 1.2$MeV. The difference is entirely due to a flatter
GEMO pattern ({\em i. e.,} more symmetric around the origin),
consistent with $A^{1/3}$ scaling. In the lighter region it is
impossible to identify common systematics, while the heavier one is
dominated by cubic parabolic trends, {\em e. g} in the EI space
between $N=82$ and 126.

If we now compare the models for the cases studied in last section we
find the following; at $t=8$ GEMO is better, at $t=16$ DZ10macro is
better, at $t=24$ and 32 both are comparable provided the filling
orders are corrected, at $t=40$ GEMO is much better. In general, the
detailed subshell structure of GEMO is a source of unphysical, jumpy,
behaviour (common to mean field formulations), while DZ10macro suffers
from distortions due to bad asymptotics; but both models are telling
the same thing about anomalous terms:
\begin{itemize}
\item Bad scaling is due to bad asymptotics. 
\item Cubic terms {\em i.e.,} three body forces are a
  fact, not a DZ10 artifact. In the next section we shall offer a
proof of the need of genuine three body forces to explain the HO-EI
transition.
\end{itemize}
 
Let us conclude with a word on the 2.5 contraction. GEMO was fitted to
single particle and single hole properties on closed shell nuclei:
they demand huge shell effects whose relative positions are little
affected by total binding energies {\em i. e.,} by correlation
effects. Once these are switched on they will produce in first
approximation parabolas of negative slope $m(D-m)$ opposite to those
in, say, Fig.~\ref{fig:Msheff}, of form $m(m-D)$. The contraction
factor simulates this mechanism. The argument is purely heuristic---we
have not specified the nature of the correlations---but conceptually
valuable as it provides an estimate of their (large) magnitude and
explains the impossibility for an effective mean field description to
describe simultaneously masses and single particle or single hole
energies on closed shells.  In DZ10 the correlation effect is directly
incorporated in a single master term that produces both shell effects
and the bulk energy of nuclear matter. A non trivial achievement.
\section{Realistic forces and the three body issue}\label{sec:3b}
The DZ strategy amounts to assume perfect potentials and proceed as if
we were solving the Schr\"odinger equation. A perfect potential should
saturate {\em i.e.,} produce the correct energies at the right
radii. It should also produce the EI closures. It does neither.
Therefore DZ postulates that the necessary mechanisms can be
incorporated into some effective master terms $M_A$ and $M_T$ and
adopts the standard shell model approach of ``freezing'' $\hbar
\omega$ at the observed radii. The present
consensus~\cite{rmp05,Sch06} demands the introduction of three body
forces, and in recent years considerable effort has been devoted to
derive them from chiral perturbation theory, with some success as
witnessed by a recent paper~\cite{Ots09} where abundant references
will be found. However, much remains to be done at the fundamental
level and here we shall follow the policy of consulting the data
searching for hints about the monopole Hamiltonian. Our first task is
to identify the form of the necessary three body terms.

NOTATIONS: in what follows we use
$X_{st}=X_s(X_t-\delta_{st})/(1+\delta_{st})$; $X^{(2)}=X(X-1)$.

We start with an isoscalar monopole two body interaction written as
$H_m=\sum_{s\le t}m_{st}V_{st}$ where the sums extend over all orbits
$st$. From now on we omit the $V_{st}$ matrix elements and consider
only the $m_{st}$ operators. Next we examine the action of $H_m$ in a
major HO shell, and call $c$ the occupied (``core'') orbits below the
major shell. It will give three types of contribution: $m_{cc'}
=D_{cc'},\; m_cm_s = D_cm_s$ and $m_{st}$, where we have replaced the
core occupancies by the degeneracies $D_c$ of the corresponding
orbits. We are left with a pure core contribution that we assume will
be taken care of by the master term, an effective single particle term
in $m_s$ and the genuine two body $m_{st}$ part.

The first key step is to adopt the ``invariant
representation''~\cite[and references therein. An early---relatively
successful but inconclusive--attempt to introduce many body
forces]{Abz91} to separate the total number operator $m_p$ in HO shell
$p$ from others that are taken to be ``orthogonal'' to it in the sense
that they vanish for $m_p=0$ and $\bar m_p=D_p-m_p=0$ {\em i.e.,} for
zero particles and zero holes. This is achieved as follows (the orbit
1 can be chosen arbitrarily),
\begin{gather}
\label{1b}
m_s\equiv m_p+\Gamma^{(1)}_{s1}\\
\label{2b}
m_{st}\equiv
\frac{1}{2}m_p^{(2)}+(m_p-1)\Gamma^{(1)}_{s1}+\Gamma^{(2)}_{st},
\end{gather}
where
\begin{gather}
\label{G1}
\Gamma_{s1}^{(1)}=\left(\frac{m_s}{D_s}-\frac{m_1}{ D_1}\right) \frac{D_sD_1}{D_s+D_1}=
-\left(\frac{\bar m_s}{ D_s}-\frac{\bar m_1}{ D_1}\right)\frac{D_sD_1}{D_s+D_1} =
-\bar\Gamma_{s1}^{(1)}\\
\label{G2}
\Gamma_{st}^{(2)}=\left(\frac{m_s^{(2)}}{ D_s^{(2)}}
+\frac{m_t^{(2)}}{ D_t^{(2)}}-\frac{2m_sm_t}{ D_sD_t}\right) \frac{D_s^{(2)}D_t^{(2)}}{(D_s+D_t)^{(2)}}=
\bar\Gamma_{st}^{(2)}
\equiv\Gamma_{st}^{(2)}(\bar m_s,\bar m_t)
\end{gather}
In reading these expressions it is convenient to fix ideas through an
example. Consider the $sd$ shell. Eq.~(\ref{1b}) gives the spectrum of
$^{17}$O. The $m$ part belongs to the master term. The single particle
energies are referred to one of them through $\Gamma_{s1}^{(1)}$
operators in Eq.~(\ref{G1}) which change sign when particles are
turned into holes in $^{39}$Ca. The normalization ensures unit
splitting between single particle states $s$ and 1.  The $m^{(2)}_p$
in Eq.~(\ref{2b}) goes with the master term. Then we have a modulation
of the single particle energies $(m_p-1)\Gamma^{(1)}_{s1}$ that makes
it possible for the splittings in $^{39}$Ca to be different from those
in $^{17}$O. Finally the $\Gamma_{st}^{(2)}$ operators which according
to Eq.~(\ref{G2}) are particle-hole symmetric and vanish for $m,\bar
m=0,1$. They reproduce the strictly two body contributions to the
centroids (average energies) of two particle and two hole
configurations. The normalizations are such as to produce splittings
of order one between the centroids.

The second key step is as metaphysical as physical: to select very few
possible operators. If too many are truly needed, our approach is
doomed. Much of the success of DZ is due to the choice of a single
$\Gamma^{(1)}_p=\Gamma^{(1)}_{j(p)r(p)}$. This may do for masses, but
a general $H_m$ needs some extra freedom. As any one body operator
referred to its centroid becomes ``strict one body'' in the sense that
it vanishes at both $m_p,\; \bar m_p=0$ two classical choices come to
mind $l\cdot s$ and $l\cdot l\equiv l(l+1)-p(p+3)/2$ (they may be
written as combinations of $\Gamma_{s1}^{(1)}$~\cite{DZ99}). For the
strict two body case the only obvious choice is
$\Gamma^{(2)}_p=\Gamma^{(2)}_{j(p)r(p)}$. For simplicity, in what
follows we refer to a single generic $\Gamma^{(1,2)}_p$.

So far we have considered one major HO shell. In general, particles
may move in different shells. The possible cross shell operators will
be of the form
$$m_pm_{p'},\;m_p\Gamma^{(1)}_{p'},\;\Gamma^{(1)}_{p}\Gamma^{(1)}_{p'}.$$
Upon introducing three body interactions we shall encounter terms of the
type
$$m_{cc'c''}\equiv D_{cc'c''},\;m_{cc'}m_s\equiv D_{cc'}m_s\; {\rm and}\;
D_cm_{st}$$ 
which will modify the two body equations~(\ref{1b},\ref{2b}). The
genuine $m_{stu}$ parts will contain $m_p,$ and $\Gamma^{(1,2)}_p$
operators plus eventually cubics $\Gamma^{(3)}_p$. By now we
can propose a list of possible operators:
\begin{gather}
  \label{eq:intra}
  {\rm intra \ shell:}\quad (a_1+b_1m_p+c_1m_p^2)\Gamma^{(1)}_p,\quad
  (a_2+b_2m_p)\Gamma^{(2)}_p,\quad a_3\Gamma^{(3)}_p\\
\label{eq:cross}
{\rm cross \ shell:}\quad
(\alpha_1m_p+\beta_1m_p^2+\gamma_1m_pm_{p'})\Gamma^{(1)}_{p'},\quad
m_p(
\alpha_2\Gamma^{(2)}_{p'}+\beta_2\Gamma^{(1)}_{p}\Gamma^{(1)}_{p'}),\quad
\alpha_3\Gamma^{(1)}_p\Gamma^{(2)}_{p'}
\end{gather}
We should not forget the pure $m$ contributions that we have decided
to ascribe to (the three body part of) the master term, see~\cite[p
437]{rmp05} for some speculation on the subject.

Everything we have said applies either to purely isoscalar operators
or to a neutron proton representation, in which case each major shell
is split in two with $m_s$ replaced by $n_s$ and $z_s$. Physically it
is preferable to work in an isospin representation involving $m_s$ and
$T_s$ operators. The reason can be understood by referring to
Fig.~\ref{fig:beld} which shows that shell effects are very much the
same for both fluids (at constant $N$ or $Z$). If we plot along lines
of constant $A$ or $T$, the former become the fairly smooth ``$\beta$
decay parabolas'' while the latter (of which we have seen many
examples) exhibit most of the shell structure, indicating its
basically isoscalar character. Let us examine how to proceed.

\subsection{The origin of the HO-EI transition}

Once we accept that three body effects are necessary, we must find
ways to identify the relevant operators. As there are several
candidates, to select them by fitting masses may give non unique
answers.  The best way to proceed---as far as the HO-EI
transition is concerned---is to look into spectroscopic information:
an apparently impossible task that turns out to be very simple.

The unequivocal sign that a strict two body treatment is not
sufficient came from exact calculations leading to a $J^{\pi}=1^+$
ground state of $^{10}$B instead of the observed $3^+$
\cite{navor,pieper}, a long standing puzzle in conventional shell
model work, where exactly the same problem exists in $^{22}$Na. Then
it was shown~\cite{Zuk03} that by changing the $V_{jj}^T,\;V_{jr}^T$
centroids of a realistic interaction R according to
\begin{equation}
  \begin{array}{l}
V_{jr}^T(\text{R})\Longrightarrow V_{jr}^T(\text{R})-(-)^T\,\kappa\\
V_{jj}^T(\text{R})\Longrightarrow V_{jj}^T(\text{R})-1.5\,\kappa \, \delta_{T,0}
  \end{array}
  \label{eq:R1}
\end{equation}
one could correct much of the spectroscopic trouble in the $p,\,sd$
and $pf$ shells. To within some fine tunings---and a major
difference---this is a time-honored two body prescription~\cite{Pas76}
that has become a common feature of effective interactions in the $pf$
shell~\cite[sections V, VB]{rmp05}. The major difference is that
$\kappa$ is a linear function of $m_p$, which makes the prescription
three body. But it (still) has a flaw: as it was originally devised to
ensure EI closures at $^{48}$Ca and $^{56}$Ni where the $V_{rr}^T$
centroids play no role, they were left out.
\begin{figure}[!ht]
\includegraphics[width=8 cm]{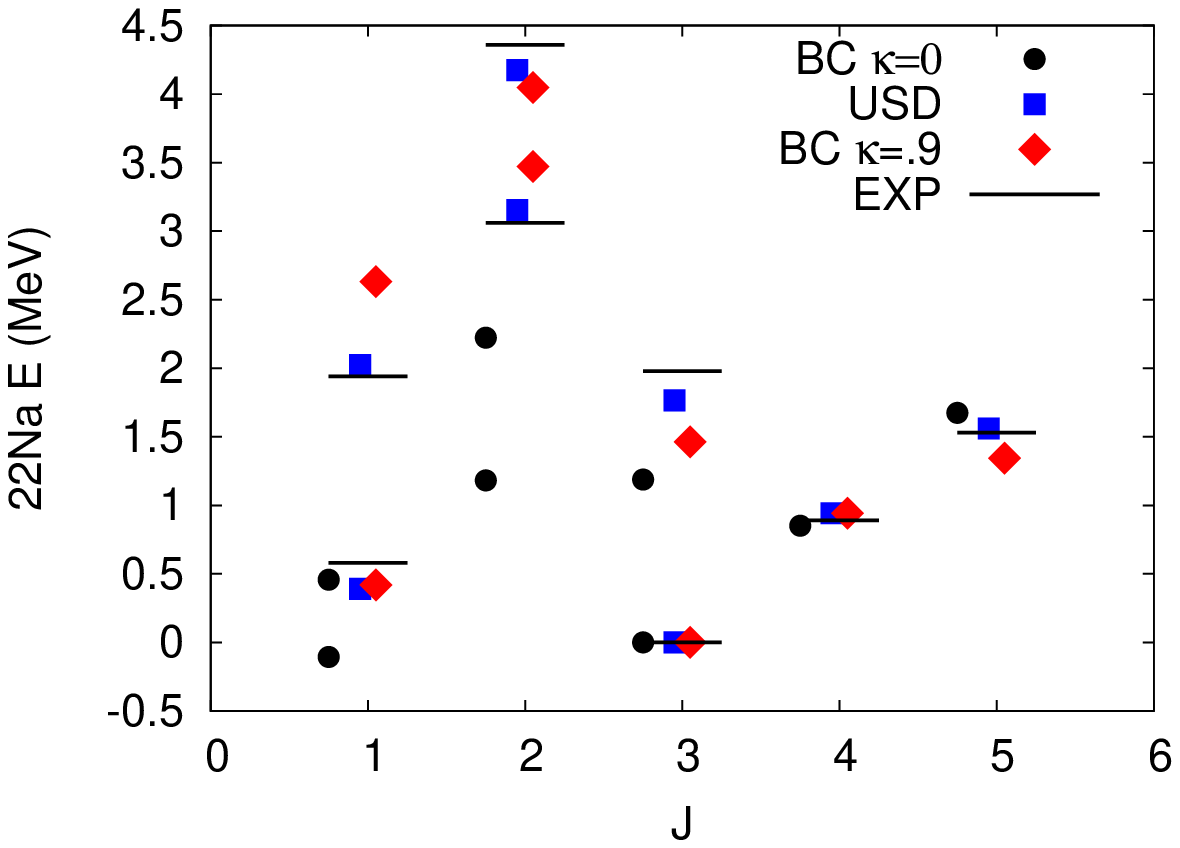} 
\includegraphics[width=8 cm]{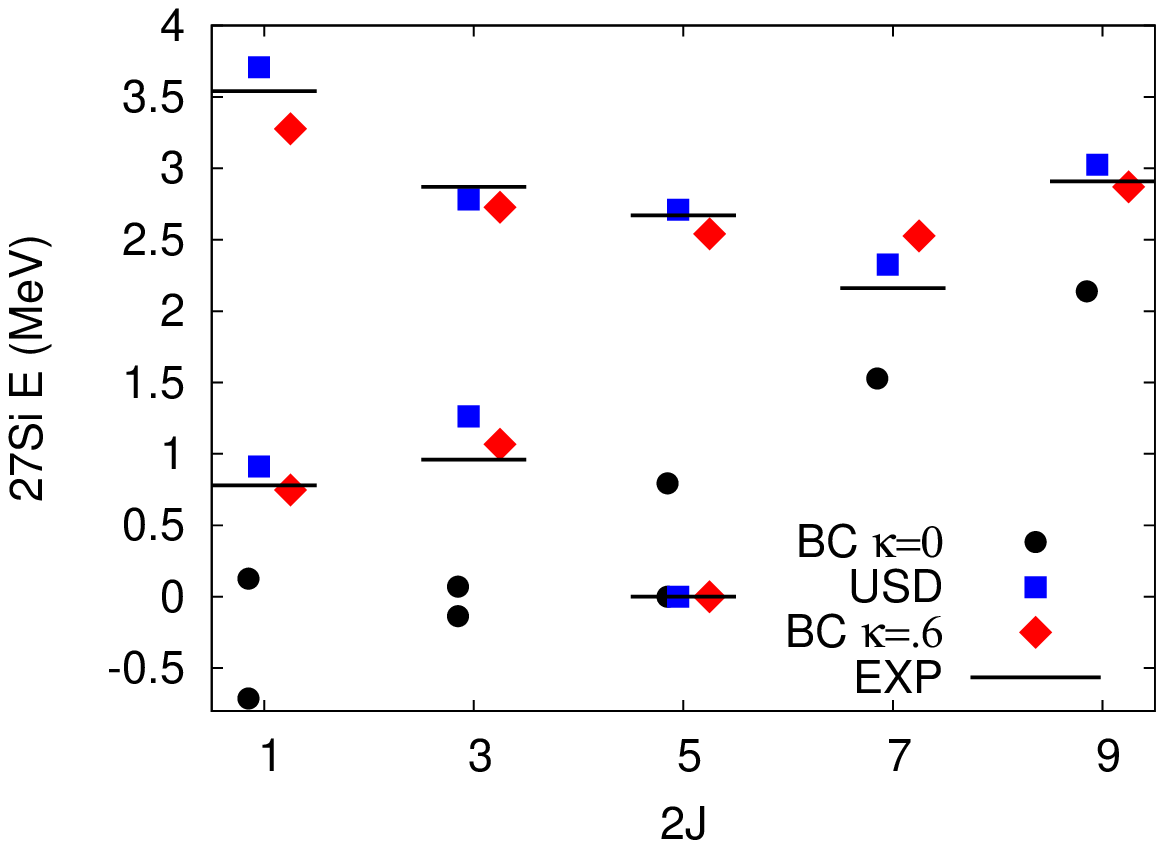} 
\caption{Spectra of $^{22}$Na and $^{27}$Si. The uncorrected and
  corrected Bonn C interaction (BC)~\cite{HJKO95} is compared with
  USD~\cite{Wi84} and experiment.}
\label{fig:Na22Si27}
\end{figure}
To include them we note that the six centroids involved, once cast as
three isoscalar and three isovector terms, reduce to four by removing
the overall $m(m-1)$ and $T(T+1)$ dependences. According to our policy
of omitting at first the isovector pieces, we are left with the terms
in Eq.~(\ref{eq:intra}). We ignore the last and redo the calculations
in~\cite{Zuk03}. We know from~\cite{Abz91} that
$(a_1+b_1m_p+c_1m_p^2)\Gamma^{(1)}_p$ plays a role, but it does not
solve the problems. Therefore we simulate its influence by pushing up
the $d_{3/2}$ orbit two MeV above its observed value (arbitrary move
but of little consequence). We are left with
$(a_2+b_2m_p)\Gamma^{(2)}_{j(p)r(p)}\equiv\kappa(m)\Gamma^{(2)}_{j(p)r(p)}$,
which is seen in Fig.~\ref{fig:Na22Si27} to work very well:
the dismal spectra produced by the original interaction (Bonn C
$\kappa=0$) become adequate in $^{22}$Na (Bonn C $\kappa(6)=0.9$) and good,
even when compared with USD, in $^{27}$Si (Bonn C $\kappa(11)=0.6$). There is
no way to find a compromise, constant, $\kappa$.

A refined approach will demand the inclusion of some other operators,
but there are reasons to believe that $\kappa(m)\Gamma^{(2)}_p$ will play
a (the) fundamental role in the HO-EI transition.

\section{Homage to Jean Duflo and conclusions}\label{sec:homage}
In all justice DZ10 should be called D10 because it was entirely the
work of the late Jean Duflo. It was designed to shed some light
on the problems of DZ28:

a) Too many parameters; b) Obscure origin of EI closures; c) Many
surface terms that lead to a change of sign at around $A=120$; d)
Anomalous $A$ scalings.

At the price of a substantial but acceptable RMSD increase, Duflo
solved a, clarified b, reduced c to a single three body term and was
left with d. He had a genius for regrouping and eliminating
theoretically plausible contributions and inventing phenomenological
ones demanded by data. Jean was a magical data manager. He was also
aware that DZ10 needed improvements, and to the last day he worked on
them. He did not find a satisfactory solution, but from his notes one
could guess that he did not worry about anomalous terms and
concentrated on basic shell formation. Our present work points to the
same direction. We have explained how the scaling anomaly can be
eliminated and demonstrated the need of genuine three body forces. It
remains to be seen whether these advances can lead to a formulation as
compact and successful as D10.

\acknowledgments
We have enjoyed some interesting exchanges at GSI and Darmstadt with
H. Feldmeier, G. Mart{\'\i}nez Pinedo, N. Pietralla and F. Thielemann.
Special thanks are due to the referee whose remarks lead to substantial improvements. This work was supported in part by Conacyt, M\'exico, by FONCICYT project 94142, by DGAPA, UNAM and by Helmholtz Alliance HA216/EMMI.

\end{document}